\documentclass[aps,10pt,prb,showpacs,twocolumn, superscriptaddress, longbibliography]{revtex4-2}
\usepackage{graphicx}
\usepackage{amssymb, amsmath}
\usepackage{amsfonts}
\usepackage{color}
\usepackage{epstopdf}
\usepackage{multirow}
\usepackage{booktabs}

\newcommand\varpm{\mathbin{\vcenter{\hbox{%
  \oalign{\hfil$\scriptstyle+$\hfil\cr
          \noalign{\kern-.3ex}
          $\scriptscriptstyle({-})$\cr}%
}}}}
\newcommand\varmp{\mathbin{\vcenter{\hbox{%
  \oalign{$\scriptstyle({+})$\cr
          \noalign{\kern-.3ex}
          \hfil$\scriptscriptstyle-$\hfil\cr}%
}}}}
\DeclareMathAlphabet      {\mathbf}{OT1}{cmr}{bx}{n}

\begin{document}

\title{Zero-point entropies of spin-jam and spin-glass states in a frustrated magnet}
\author{C.~Piyakulworawat}
\affiliation{Department of Physics, Faculty of Science, Mahidol University, Bangkok, 10400, Thailand}
\author{A.~Thennakoon}
\affiliation{Department of Physics, University of Virginia, Charlottesville, Virginia, 22904, USA}
\author{J.~Yang}
\affiliation{Department of Physics, University of Virginia, Charlottesville, Virginia, 22904, USA}
\affiliation{Department of Physics, New Jersey Institute of Technology, Newark, New Jersey, 07102, USA}
\author{H.~Yoshizawa}
\affiliation{Neutron Science Laboratory, Institute for Solid-State Physics, The University of Tokyo, Kashiwa, Chiba, 277-8581, Japan}
\author{D.~Ueta}
\affiliation{Neutron Science Laboratory, Institute for Solid-State Physics, The University of Tokyo, Kashiwa, Chiba, 277-8581, Japan}
\author{T.~J~Sato}
\affiliation{Institute of Multidisciplinary Research of Advanced Materials, Tohoku University, Sendai, Miyagi, 980-8577, Japan}
\author{K.~Sheng}
\affiliation{Center for Condensed Matter Sciences, National Taiwan University, Taipei, 10617, Taiwan}
\affiliation{Taiwan Consortium of Emergent Crystalline Materials, National Science and Technology Council, Taipei, 10622, Taiwan}
\author{W.-T.~Chen}
\affiliation{Center for Condensed Matter Sciences, National Taiwan University, Taipei, 10617, Taiwan}
\affiliation{Taiwan Consortium of Emergent Crystalline Materials, National Science and Technology Council, Taipei, 10622, Taiwan}
\author{W.-W.~Pai}
\affiliation{Center for Condensed Matter Sciences, National Taiwan University, Taipei, 10617, Taiwan}
\affiliation{Taiwan Consortium of Emergent Crystalline Materials, National Science and Technology Council, Taipei, 10622, Taiwan}
\author{K.~Matan}
\email[Corresponding author: ]{kittiwit.mat@mahidol.ac.th}
\affiliation{Department of Physics, Faculty of Science, Mahidol University, Bangkok, 10400, Thailand}
\affiliation{ThEP, Commission of Higher Education, Bangkok, 10400, Thailand}
\author{S.-H.~Lee}
\email[Corresponding author: ]{sl5eb@virginia.edu}
\affiliation{Department of Physics, University of Virginia, Charlottesville, Virginia, 22904, USA}

\date{\today}
\begin{abstract}
Thermodynamics studies of a prototypical quasi-two-dimensional frustrated magnet Ba$_2$Sn$_2$ZnCr$_{7p}$Ga$_{10-7p}$O$_{22}$ where the magnetic Cr$^{3+}$ ions are arranged in a triangular network of bipyramids show that the magnetic zero-point entropy for $p=0.98$ is 55(1)\% of the entropy expected when the Cr$^{3+}$ moments are fully disordered. Furthermore, when combined with a previous neutron scattering study and the perimeter scaling entropy of a spin jam, the analysis reveals that with decreasing $p$, i.e., doping of the nonmagnetic Ga$^{3+}$ ions, the variation in the magnetic zero-point entropy can be well explained by the combined effects of the zero-point entropy of the spin jam state and that of weakly coupled orphan spins, shedding light on the coexistence of the two types of spin states in quantum magnetism.
\end{abstract}
\maketitle

\section{Introduction}\label{sec1}


Zero-point entropy, i.e., the entropy at absolute zero temperature, of a macroscopic system has been a strenuously debated topic ever since the introduction of the third law of thermodynamics. One example of magnetic solids that could possess a finite zero-point entropy is the so-called spin glasses. The spin-glass state can exist in dilute magnetic alloys in which nonmagnetic metals are doped with magnetic ions at low concentrations.  These magnetic impurities can interact with one another through the Ruderman-Kittel-Kasuya-Yosida (RKKY) interaction. Below the spin-glass transition temperature, the magnetic moments of impurities freeze in random directions without long-range ordering due to the randomness of the RKKY interactions, resulting in a finite zero-point entropy. The zero-point entropy in spin glasses has been estimated theoretically by Edwards and Tanaka, who predicted the values for long-range-interacting Ising and $XY$ spin glasses to be 1.66 and 4.30 Jmol$^{-1}$K$^{-1}$, respectively \cite{Edwards1980, Tanaka1980}. Experimentally, the zero-point entropy of a dilute dipolar-coupled Ising spin glass LiHo$_p$Y$_{1-p}$F$_4$ with $p = 0.167$ was measured and found to be close to 1.66 Jmol$^{-1}$K$^{-1}$, consistent with the theoretical prediction \cite{Quilliam2007}.

An interesting question that arises is what will happen to the zero-point entropy if, unlike in the dilute magnetic alloys, the magnetic ions are densely populated and strongly interact with each other. The so-called geometrically frustrated magnets are the case in point. For example, pyrochlore rare-earth oxides $A_2B_2$O$_7$ which exhibit the so-called spin-ice state at low temperatures have similar degenerate ground-state configurations to water ice in which two spins must point inwards while the other two point out of the tetrahedron \cite{Bramwell2001}. Surprisingly, the zero-point entropies of the Ho$_2$Ti$_2$O$_7$ and Dy$_2$Ti$_2$O$_7$ spin ices have been reported to exhibit a value close to that of water ice \cite{Ramirez1999, Higashinaka2003, Lau2006, Zhou2007}. CuAl$_2$O$_4$ and CuGa$_2$O$_4$ spinels with magnetic ions residing in the diamond sublattice have also been found to manifest a finite zero-point entropy \cite{Fenner2009, Nirmala2017}. Other frustrated lattices can have local zero-energy modes in the mean-field level, i.e., the weathervane modes in the two-dimensional kagome antiferromagnets \cite{Harris1992, Chalker1992, Chubukov1992, Sachdev1992} and the antiferromagnetic hexagon modes in the three-dimensional spinel ZnCr$_2$O$_4$ \cite{Lee2002}, which can induce macroscopic ground-state degeneracy and thus a finite zero-point entropy.

These densely populated geometrically frustrated magnets can exhibit a magnetic glassy state at low temperatures that is called a spin jam \cite{Klich2014}. While the canonical spin-glass state arises due to the random RKKY interactions, the spin-jam state can arise from quantum fluctuations \cite{Klich2014}. The essential distinction between the two glassy states is in their energy landscape topology. Quantum fluctuations render the energy landscape of the spin jam to be non-hierarchical and have a flat but rugged shape. On the contrary, for the spin glass, the energy landscape is hierarchical and has a rugged funnel shape \cite{Samarakoon2016}. To date, the crossover between these glassy states has been observed in the dynamic susceptibility measurements and the memory effects when the spin density is varied in the systems \cite{Yang2015, Samarakoon2017}.  In this letter, we report experimental evidence of the zero-point entropy in the glassy state of a QS ferrite-derived compound Ba$_2$Sn$_2$ZnCr$_{7p}$Ga$_{10-7p}$O$_{22}$ (BSZCGO) \cite{Hagemann2001, Bono2004_1, Bono2004_3, Mutka2006}, a realization of the frustrated triangular network of bipyramids. The material can be viewed as a stacking of two types of blocks, the nonmagnetic `Q' block and the magnetic `S' block, alternating with each other.  Furthermore, through our analysis, we show how the spin-jam state crosses over to the spin-glass state as the spin density $p$ varies in terms of the low-lying excitations and the zero-point entropy using DC magnetic susceptibility and heat capacity measurements down to 0.5~K.

\begin{figure}
    \centering
    \includegraphics[width=\linewidth]{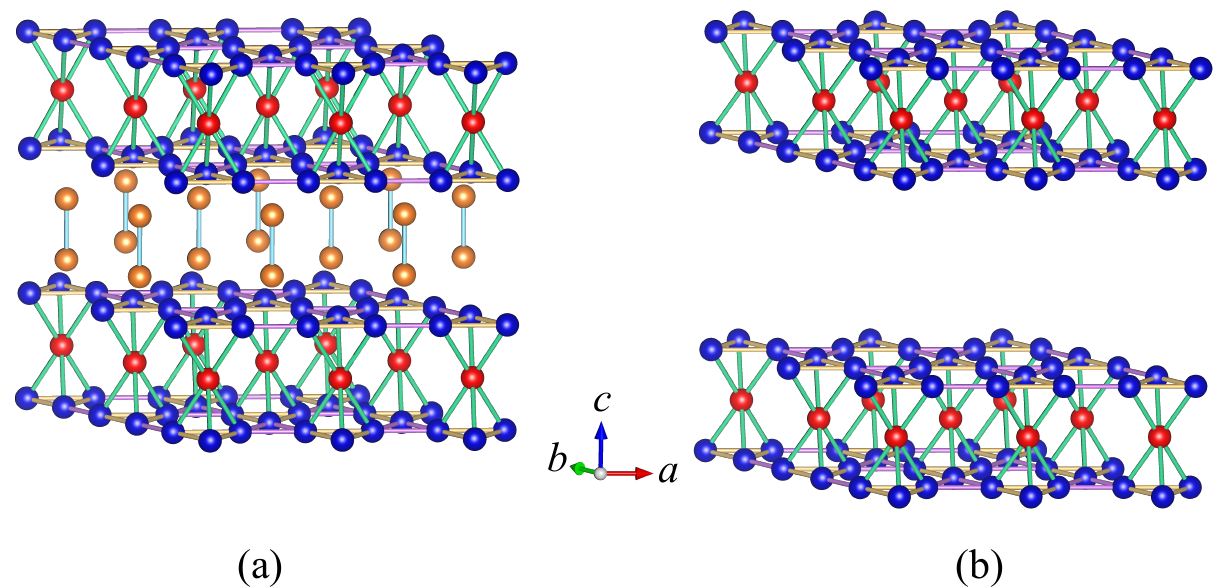}
    \caption{Magnetic lattices of SCGO and BSZCGO. A triangular network of bipyramids consists of two kagome layers (blue spheres) sandwiching an intermediate triangular layer (red spheres). Bonds shown in different colors have different lengths. (a) In SCGO, a triangular network of dimers (orange spheres) separates the successive pyrochlore slabs. (b) In BSZCGO, successive pyrochlore slabs are well separated, and there are no Cr\textsuperscript{3+} ions in between. Axes represent the crystallographic axes of the lattices.}
    \label{fig1}
\end{figure}

Since its discovery, SrCr$_{9p}$Ga$_{12-9p}$O$_{19}$ (SCGO), a cousin compound to BSZCGO, has been a good model system for the triangular network of bipyramids or pyrochlore slab [see Fig. \ref{fig1}(a)] \cite{Broholm1990, Ramirez1992, Martinez1992, Lee1996, Mondelli1999, Limot2002, Iida2012, Yang2015}. The system, however, has triangular layers of spin dimers formed by Cr$^{3+}$ spins [orange spheres in Fig. \ref{fig1}(a)], residing between the pyrochlore slabs \cite{Lee1996}. The existence of the extra magnetic layers of dimers complicates the physics of the pure pyrochlore slab. BSZCGO, on the other hand, does not comprise the spin dimer layers. The crystal structure of BSZCGO is characterized by the hexagonal system with the space group $P\bar{3}m1$ and lattice parameters $a$ = $b$ = 5.8568(1)~\AA~and $c$ = 14.2537(3)~\AA~for the sample with $p$ = 0.97 \cite{Bonnet2004}. The magnetic $s = \frac{3}{2}$ Cr$^{3+}$ ions form the pyrochlore slabs, and the successive slabs are separated by about 10~\AA, making the pyrochlore slabs well isolated and quasi-two-dimensional [see Fig. \ref{fig1}(b)]. There are, however, two types of intrinsic disorder in BSZCGO. Firstly, nonmagnetic Ga$^{3+}$ ions inevitably share $6i$ and $1a$ sites with Cr$^{3+}$ ions leading to the highest possible value of the spin density $p$ to be about 0.97 \cite{Mutka2007}. Furthermore, Ga$^{3+}$ ions also share the $2d$ site with Zn$^{2+}$ ions in a 1:1 ratio which causes structural strains and in turn renders bond disorders between Cr$^{3+}$ ions \cite{Bono2004_1, Bono2004_2}. Despite these disorders, BSZCGO is the best model system to explore the physics of frustration in the triangular network of bipyramids due to its robustness against small disorders \cite{Klich2014, Syzranov2022}. BSZCGO exhibits a freezing transition with $T_\mathrm{f}$ around 1.5 K for $p$ = 0.97 \cite{Hagemann2001, Bono2005}. In this clean limit, $p \to 1$, the magnetic heat capacity $C_\mathrm{mag}$ has been observed to show a $T^2$ dependence below $T_\mathrm{f}$ \cite{Hagemann2001}, indicative of the unconventional glassy state.

\section{Experimental details}

Ten powder samples of BSZCGO with $0.44 \le p \le 0.98$ and a nonmagnetic sample with $p = 0$ (Ba$_2$Sn$_2$ZnGa$_{10}$O$_{22}$) were prepared with standard solid-state reactions. A stoichiometric mixture of BaCO$_3$, SnO$_2$, ZnO, Ga$_2$O$_3$, and Cr$_2$O$_3$ were intimately ground and pelleted. The pellet was put in an alumina crucible and sintered in air at 1400 °C for 48 hours with intermediate grinding. X-ray diffraction was performed at room temperature for each sample to verify the crystal structure and to determine Cr$^{3+}$ concentration within the sample (see Section III in \cite{SM} for details).  The temperature dependence of the DC magnetic susceptibility was measured using a commercial SQUID magnetometer from 0.5 K up to 20 K with an applied magnetic field of 0.01 T. The measurements were done with both field-cooled (FC) and zero-field-cooled (ZFC) methods. Susceptibility data of samples with $p < 0.67$ were not taken as these samples have transition temperatures lower than 0.5 K. The temperature dependence of molar heat capacity was measured with a commercial physical property measuring system utilizing a thermal relaxation technique. Pelleted powder samples ranging in mass from 1 to 7 mg were affixed using Apiezon grease to a platform equipped with a heater and thermometer. The molar heat capacity from 0.5 K to 10 K was measured with the $^3$He option and from 3 K to 50 K (up to room temperature for the $p=0.98$ and $p=0$ samples) with the $^4$He option in a zero magnetic field (see Figs. S1 and S3 in \cite{SM} for all raw heat capacity data). As shown in Fig. S1, above $\sim$50 K, the molar heat capacity of the $p=0.98$ and $p=0$ samples coincides with one another indicative of vanishing magnetic contribution to the magnetic sample above such temperature. The magnetic heat capacity was obtained by subtracting the interpolated molar heat capacity of the nonmagnetic ($p=0$) sample from that of the magnetic ($p\neq0$) samples without artificially rescaling the high-temperature data.

\section{Results and discussion}\label{sec3}
\subsection{Low-lying excitations}

\begin{figure}
    \centering
    \includegraphics[width=\linewidth]{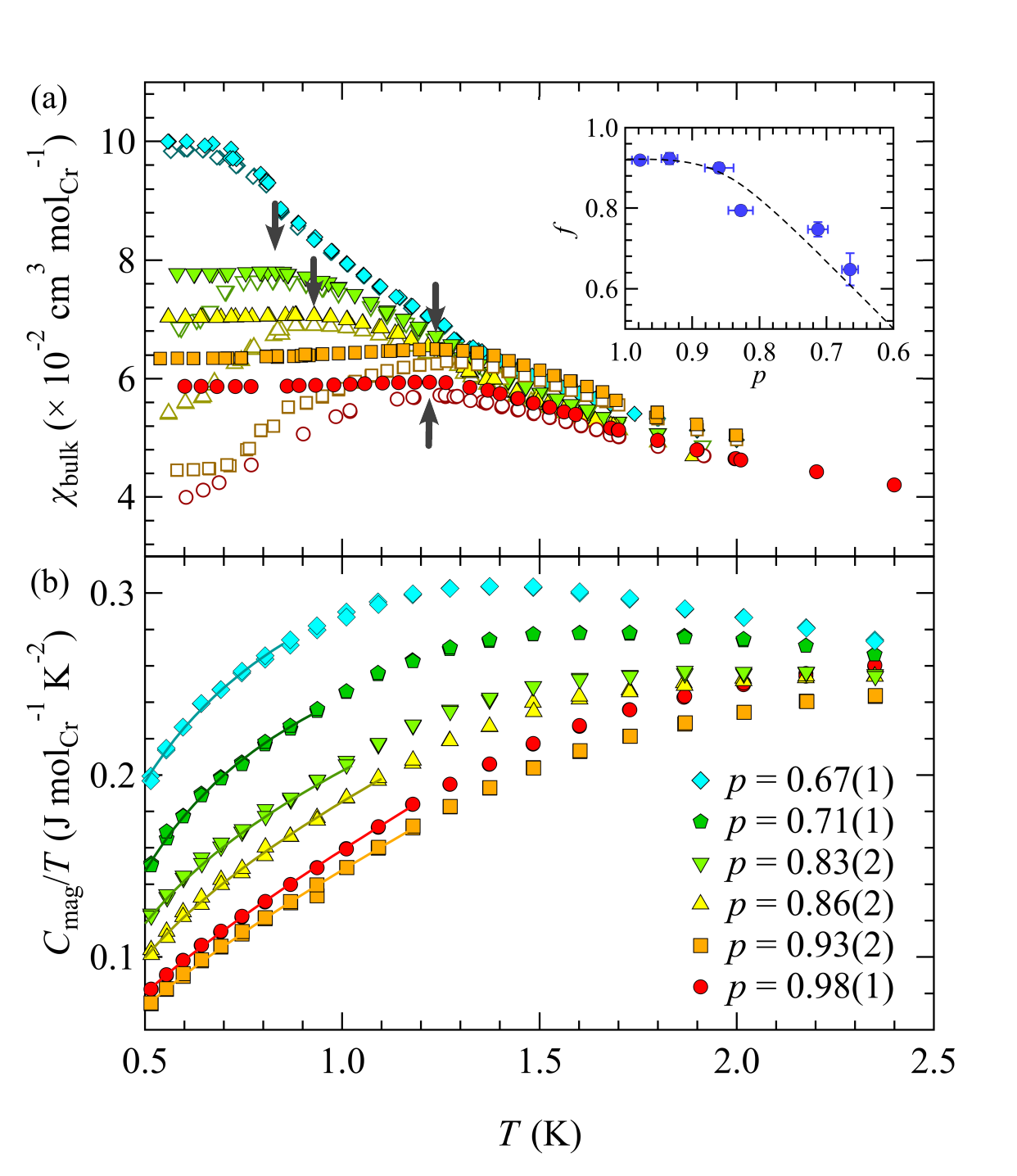}
    \caption{The \textit{T} dependence of DC magnetic susceptibility and magnetic heat capacity. (a) The DC magnetic susceptibility in the temperature range covering the freezing transition of samples with \textit{p} $\ge$ 0.67.  Open symbols represent ZFC data. Arrows mark \textit{T}\textsubscript{f} for each sample. The inset shows \textit{f} as a function of \textit{p}, where \textit{f} is the fractional population of the spin-jam state. The dashed line in the inset is a guiding line. (b) \textit{C}\textsubscript{mag}/\textit{T} data at low temperatures. Solid lines are best fits to the two-state model with fitting parameters summarized in Table \ref{table1}.}
    \label{fig2}
\end{figure}

Figure \ref{fig2}(a) shows the DC magnetic susceptibility data of five samples that exhibit the freezing transitions at $T_\mathrm{f}$ indicated by the bifurcation of FC and ZFC data. As shown in Fig. \ref{fig2}(a) (see also Fig. \ref{fig4}(a) for $T_\mathrm{f}$ obtained from magnetic heat capacity data), $T_\mathrm{f}$ is found to decrease with increasing vacancy density (decreasing $p$), which is consistent with the spin-jam theory \cite{Klich2014, Yang2015}. Note that for canonical spin glasses, the impurity dependence of $T_\mathrm{f}$ behaves differently; $T_\mathrm{f}$ increases with increasing magnetic impurity density \cite{Maletta1979, Cannella1972, Nagata1979}. The nature of the glassy states can be studied more carefully via the behavior of the $T$-dependent magnetic heat capacity $C_\mathrm{mag}$. Figure \ref{fig2}(b) shows $C_\mathrm{mag}/T$ as a function of $T$ in the low-temperature region of six samples with $p \ge 0.67$ (see Fig. S4 in \cite{SM} for the low-temperature data of all samples). For $p \ge 0.93$, $C_\mathrm{mag}$ exhibits a clear quadratic $T^2$-dependence. On the other hand, for $p \le 0.86$, $C_\mathrm{mag}$ begins to deviate from the quadratic behavior. To quantitatively analyze the data, we assume that the thermodynamics of the spin fluctuations can be characterized by two modes; one is the hydrodynamic Halperin-Saslow (HS) mode \cite{Halperin1977} that is a characteristic of the spin-jam state and yields $C_\mathrm{HS} = AT^2 + B$ for a two-dimensional system \cite{Klich2014}, where $B$ is a temperature-independent term \cite{Ramirez1992, Podolsky2009}, and the other is the localized two-level (TL) system due to spin-glass clusters generated by the non-magnetic doping and yields $C_\mathrm{TL} \propto T$ \cite{Anderson1971, Yang2015}. The coefficient $A$ of the $T^2$ term in $C_\mathrm{HS}$ is inversely proportional to the spin wave velocity squared $v^2$ for a two-dimensional system,
\begin{align}
    A = \frac{9\zeta(3)k_\mathrm{B}^2V_\mathrm{c}R}{\pi\hbar v^2d},
    \label{A}
\end{align}
where $\zeta$ is the Riemann zeta function, $k_\mathrm{B}$ is Boltzmann's constant, $V_\mathrm{c}$ is the unit cell volume, and $d$ is the spacing of successive bilayers \cite{Halperin1977, Podolsky2009}.

Since the population ratio of the spin jam to the spin glass clusters can vary with the spin density $p$, we have fitted the magnetic heat capacity of each sample to the following phenomenological formula, $C_\mathrm{mag}=fC_\mathrm{HS}+(1-f)C_\mathrm{TL}$, where $f$ and $1-f$ are the fraction of the spin-jam state and that of the spin-glass state, respectively. The fitting range of $T$ is from the base temperature of 0.5 K to about $T_\textrm{f}$. As shown by the solid lines in Fig. \ref{fig2}(b), the phenomenological formula fits the data well for $p \ge 0.67$ while the data for $p < 0.67$ could not be fitted due to the lack of enough data points below $T_\mathrm{f}$. The fitted parameters are summarized in Table \ref{table1}. As the spin density $p$ decreases below 0.93, $f$ gradually decreases roughly linearly as shown in the inset of Fig.~\ref{fig2}(a). In other words,  the glassy state of BSZCGO continually crosses over from a dominantly spin-jam state to a mixed state with a considerable spin-glass state as $p$ decreases.

As another quantitative verification of the HS modes being dominant for large values of $p$, an energy scale associated with this mode can be estimated from the coefficient $A$ of the quadratic term of $C_\mathrm{HS}$. The spin stiffness $\rho_\mathrm{s}$ and the spin wave velocity $v$ are related by $v = \gamma\sqrt{\rho_\mathrm{s}/\chi}$, where $\chi$ is the magnetic susceptibility and $\gamma$ is the gyromagnetic ratio, and $A$ is related to $v$ via Eq. \ref{A}. From the spin stiffness $\rho_\mathrm{s}$, the HS energy $E_\mathrm{HS}$ is expressed as
\begin{align}
    \frac{E_\mathrm{HS}}{k_\mathrm{B}} = \frac{9\zeta(3)}{\pi}\frac{k_\mathrm{B}^2}{g^2\mu_\mathrm{B}^2}\frac{\chi}{A},
    \label{E_HS}
\end{align}
where $g$ is the Landé factor and $\mu_\mathrm{B}$ the Bohr magneton \cite{Podolsky2009}. The magnetic susceptibility $\chi$ is obtained from the measured susceptibility below $T_\mathrm{f}$. As shown in Table \ref{table1}, this formula yields $E_\mathrm{HS}$ that is comparable to the freezing temperatures for the two samples with the highest spin densities $p$, which supports our interpretation of the dominant glassy state for $p \ge 0.93$ being the spin jam. For $p < 0.93$, the spin-glass population starts to grow, and its susceptibility contributes significantly to the measured value, resulting in the overestimation of $\chi$ used in Eq. \ref{E_HS} and hence the overestimation of $E_\mathrm{HS}$ for $p < 0.93$.

\begin{table*}
    \centering
    \caption{Fitting parameters of the Halperin-Saslow modes in BSZCGO where \textit{p} is the spin density obtained from the X-ray diffraction measurements (see Section III in \cite{SM}), \textit{f} is the spin-jam population fraction, \textit{T}\textsubscript{f, $\chi$} and \textit{T}\textsubscript{f, $C_\textrm{mag}$} are the freezing temperatures extracted from the magnetic susceptibility and heat capacity, respectively, \textit{A} is the coefficient of the quadratic term of \textit{C}\textsubscript{HS}, and \textit{E}\textsubscript{HS}/\textit{k}\textsubscript{B} is the energy scale of the HS modes. Numbers in parentheses represent errors. The values of \textit{A} for the last two samples have errors larger than itself.}
    \begin{tabular}{cccccc}
    \hline
    \hline
    ~~~~~~\textit{p}~~~~~~ & ~~\textit{f} ~~& ~~~\textit{T}\textsubscript{f, $\chi$}  (K)~~~ & ~~~\textit{T}\textsubscript{f, $C_\textrm{mag}$}  (K)~~~ & ~\textit{A}~ (Jmol\textsubscript{Cr}\textsuperscript{-1}K\textsuperscript{-3})~~~ & \textit{E}\textsubscript{HS}/\textit{k}\textsubscript{B}  (K)  ~~~\\
    \hline
    ~0.98(1)~ & ~0.92(1)~ & 1.22(5) & 1.18(9) & 0.130(2) & 0.9(1) \\
    0.93(2) & 0.92(1) & 1.24(5) & 1.18(9) & 0.120(5) & 1.0(1) \\
    0.86(2) & 0.90(1) & 0.93(5) & 1.09(8) & 0.10(1)  & 1.3(1) \\
    0.83(2) & 0.79(2) & 0.83(5) & 1.01(8) & 0.08(1)  & 1.8(2) \\
    0.71(1) & 0.75(2) &    -    & 0.94(7)    &     -    &    -   \\
    0.67(1) & 0.65(4) &    -    &  0.87(7)   &     -    &    -   \\
    \hline
    \hline
    \end{tabular}
    \label{table1}
\end{table*}

\subsection{Zero-point entropies}

The evolution of the glassy states as a function of spin density $p$ may also be investigated in terms of entropy. In general, upon cooling, a magnetic system gradually releases its magnetic entropy, and an ordinary magnet releases all of its magnetic entropy when the system exhibits long-range order below the ordering temperature. On the other hand, disordered magnets would not release all their magnetic entropy due to strong frustrations, giving rise to finite zero-point entropy. Also, it should be emphasized that the spin-jam and spin-glass states may have different characteristic entropies. 

\begin{figure}
    \centering
    \includegraphics[width=\linewidth]{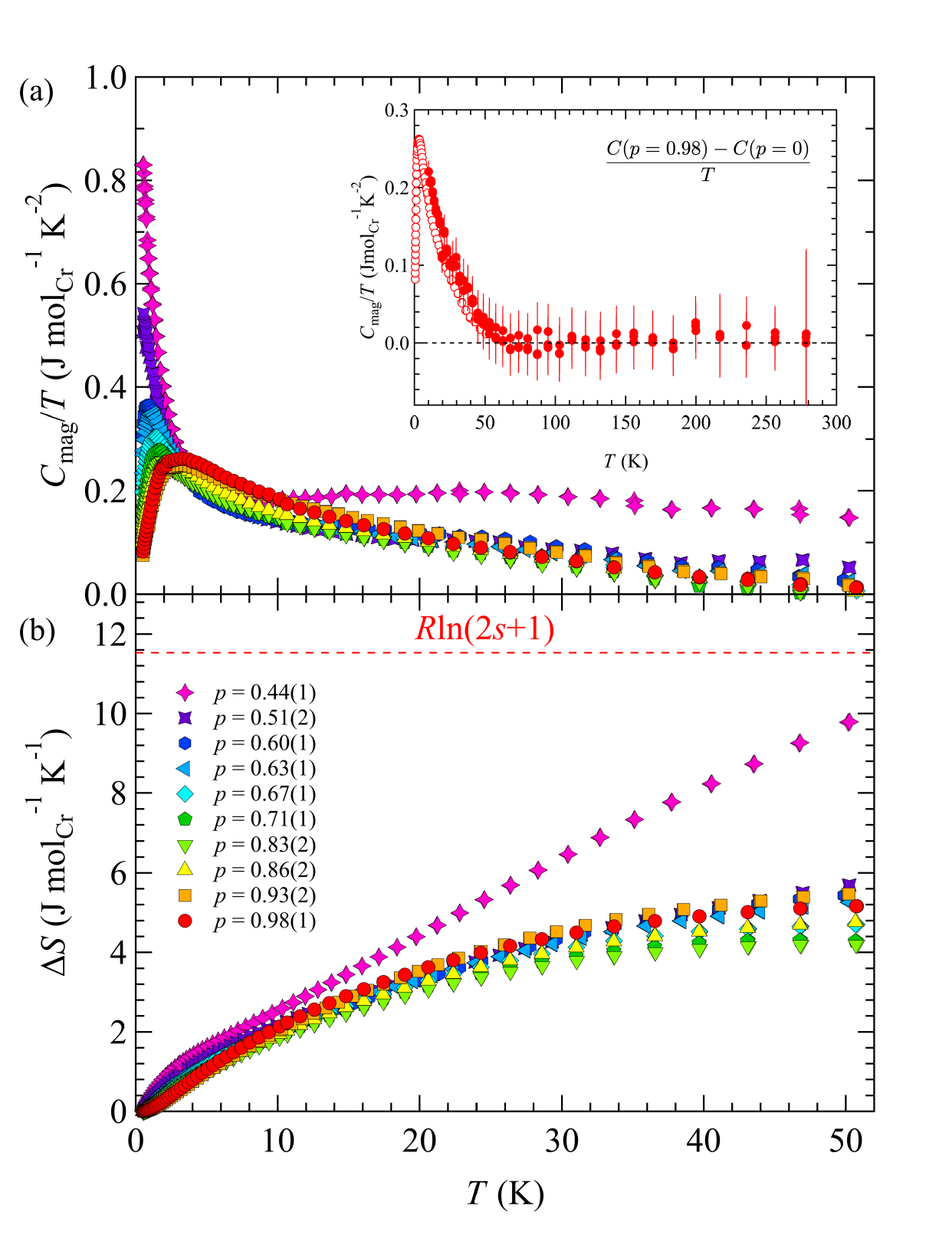}
    \caption{The \textit{T} dependence of \textit{C}\textsubscript{mag}/\textit{T} and $\Delta$\textit{S}(\textit{T}). (a) \textit{C}\textsubscript{mag}/\textit{T} for all samples up to 50 K. The inset shows \textit{C}\textsubscript{mag}/\textit{T} for the $p=0.98$ sample up to room temperature. Open symbols denote the data from the low-temperature measurements (up to 50 K)  as shown in the main panel, while closed symbols represent the data from the high-temperature measurements (up to room temperature).  (b) $\Delta$\textit{S}(\textit{T}) for all samples obtained by integrating \textit{C}\textsubscript{mag}/\textit{T} data. The red dashed line indicates \textit{S}\textsubscript{max} = \textit{R}$\ln4$.}
    \label{fig3}
\end{figure}

Entropy can be estimated from the heat capacity data as 

\begin{align}
    S(T_\mathrm{base}, T) = S_0 + \Delta S(T_\mathrm{base}, T) = S_0 + \int^{T}_{T_\mathrm{base}}\frac{C_\mathrm{mag}}{T}dT,
    \label{delta_S}
\end{align}
where $S_0$ is the zero-point entropy and $T_\mathrm{base}$ is the base temperature of 0.5 K. Thus, by investigating how $\Delta S$ evolves with increasing $p$ we can study how the zero-point entropy $S_0$, i.e., the entropy of the glassy state, evolves. From the $C_\mathrm{mag}/T$ data shown in Fig. \ref{fig3}(a), we numerically calculated and plotted $\Delta S(T)$ in Fig. \ref{fig3}(b). To confirm that the magnetic heat capacity tends to zero above 50 K, the heat capacity of the $p\sim0.98$ and $p=0$ samples were measured up to room temperature (see Fig. S1 in \cite{SM} for the raw data). At high temperatures above 50 K, the heat capacity data of the magnetic ($p\sim0.98$) and nonmagnetic ($p=0$) samples coincide with each other, indicating that the magnetic contribution to the heat capacity is negligible as shown in the inset of Fig. \ref{fig3}(a). The result leads to the conclusion that there is no further increase in the magnetic entropy at high temperatures. The magnetic entropy change present below 0.5 K, $\Delta S(T < 0.5\;\mathrm{K})$, can be approximated by linear extrapolation down to absolute zero temperature. Considering the $p = 0.60$ sample, we obtained $\Delta S(T < 0.5\;\mathrm{K})$ equal to about 0.07 Jmol$_\mathrm{Cr}^{-1}$K$^{-1}$ which is only $\sim$1\% of $\Delta S(50\;\mathrm{K})$.  This value decreases for higher $p$, and is smaller than the error bar of $\Delta S(50\;\mathrm{K})$, which is $\sim$0.4 Jmol$_\mathrm{Cr}^{-1}$K$^{-1}$. Hence, we conclude that the presence of $\Delta S$ below 0.5 K is insignificant and can be safely ignored. The shortfall of entropy is thus attributed to the zero-point entropy. We note that the magnetic Cr$^{3+}$ ion has spin $s = \frac{3}{2}$ with the expected maximum magnetic entropy of $R\ln(2s+1)$ = 11.53 Jmol$_\mathrm{Cr}^{-1}$K$^{-1}$, which is represented by the horizontal red dashed line in Fig. \ref{fig3}(b). Here, mol$_\mathrm{Cr}$ represents the unit of moles of Cr$^{3+}$ ions.

As shown in Fig. \ref{fig3}(b), for $p > p_\mathrm{c} = 0.5$, where $p_\mathrm{c}$ is the percolation threshold for the magnetic lattice \cite{Henley2001}, the entropy released, $\Delta S$(0.5 K, 50 K), between 0.5 K ($< T_\mathrm{f}$) and 50 K ($\gg T_\mathrm{f}$) is only about half of the maximum magnetic entropy $S_\mathrm{max}$. For instance, $\Delta S$(0.5 K, 50 K) $= 0.45(1) S_\mathrm{max}$ for BSZCGO(0.98). This result indicates that the entropy that is not released down to 0.5 K is extensive; $S_0(p=0.98) = 0.55(1)S_\mathrm{max} = $ 6.34(14) Jmol$_\mathrm{Cr}^{-1}$K$^{-1}$. On the other hand, for $p = 0.44 < p_\mathrm{c}$, $S(T_\mathrm{base}, T)$ at 50 K is close to $S_\mathrm{max}$, $\Delta S$(0.5 K, 50 K) $\approx$ 0.8$S_\mathrm{max}$. This result implies that the extensive zero-point entropy for $p > p_\mathrm{c}$ is due to the collective frustrated interactions in the quasi-two-dimensional triangular network of bipyramids. A similar observation was reported for SCGO(0.89) in which at 100 K the magnetic entropy is recovered by only 52\% \cite{Ramirez2000}. 

\begin{figure}
    \centering
    \includegraphics[width=\linewidth]{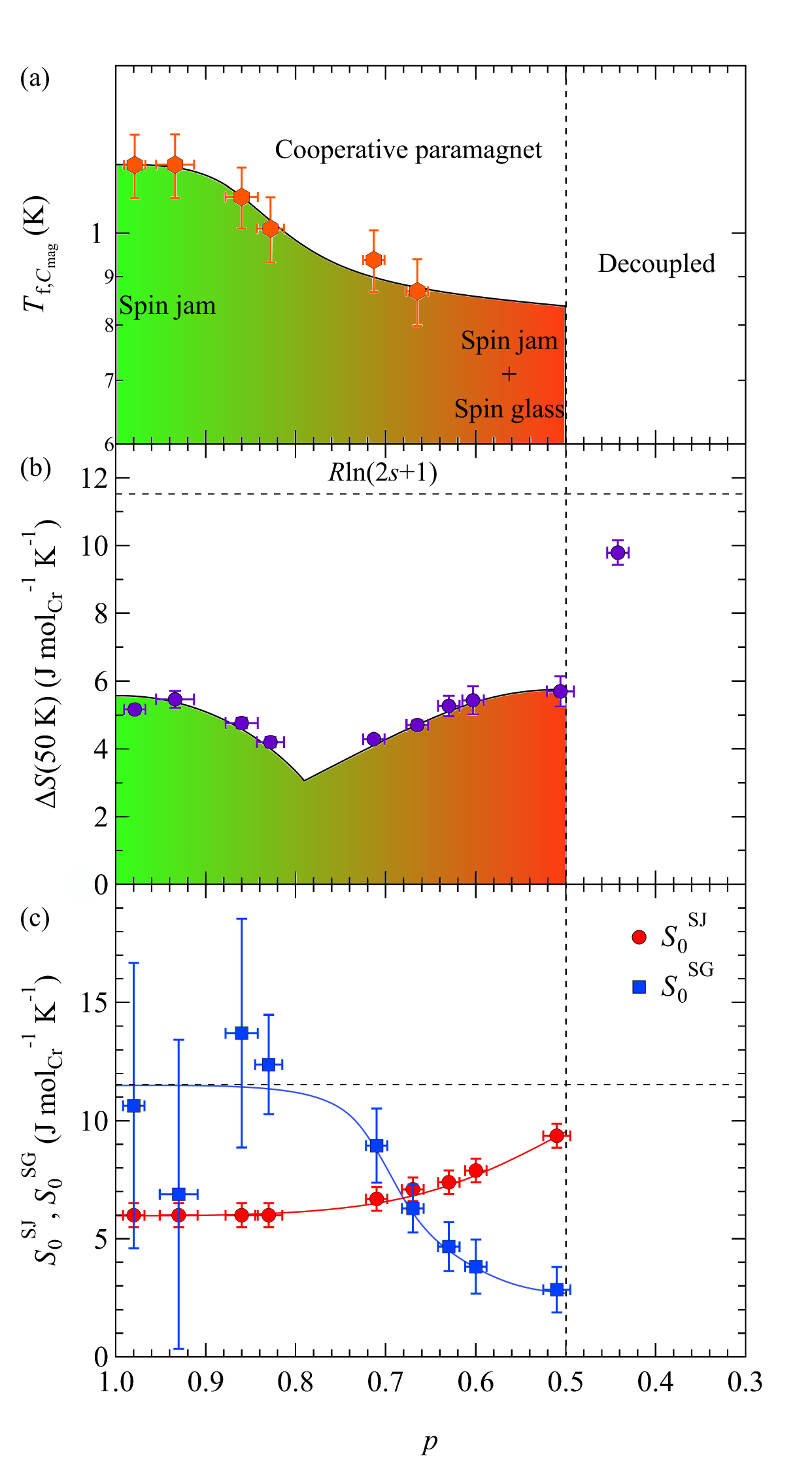}
    \caption{The $p$ dependence of (a) $T_{\textrm{f},C_\mathrm{mag}}(p)$ and (b) $\Delta S(p)$ between 0.5 and 50 K. The horizontal dashed line is $S_\textrm{max} = R\ln4 = 11.53$ Jmol\textsubscript{Cr}\textsuperscript{-1}K\textsuperscript{-1}. The vertical dashed line indicates the percolation threshold $p_\textrm{c}=0.5$. The solid lines are guides to the eyes. The gradient color represents the crossover of the system from spin-jam to spin-glass state. (c) The \textit{p} dependence of the zero-point entropy of the spin jam $S_0^\mathrm{SJ}(p)$ and that of the spin glass $S_0^\mathrm{SG}(p)$  are calculated using Eq.~\ref{S_0}, where $S_0^\textrm{tot}(p)$  is obtained from $S_0^\textrm{tot}(p) = S_\textrm{max}-\Delta S(p)$.}
    \label{fig4}
\end{figure}

A close examination of $\Delta S$(0.5 K, 50 K) as a function of $p$ reveals an interesting dependence on $p$. As shown in Fig. \ref{fig4}(b), as $p$ decreases from 0.98 to 0.83, $\Delta S$(0.5 K, 50 K) decreases by $\sim$25\%. As a result, the zero-point entropy $S_0$ increases as $p$ decreases from 0.98 to 0.83. Upon further decreasing $p$ below 0.71, $\Delta S$(0.5 K, 50 K) increases again, i.e., $S_0$ decreases. To understand the dip in $\Delta S$(0.5 K, 50 K) as a function of $p$, we first note that our analysis of the $T$-dependent $C_\mathrm{mag}$ data indicates that both spin-glass and spin-jam clusters coexist and their fraction changes with the spin density $p$. This implies that, since the spin-jam and the spin-glass states are expected to have different zero-point entropies, the measured total zero-point entropy $S_0^\mathrm{tot}$ have to include both contributions, the zero-point entropy of spin jam $S_0^\mathrm{SJ}$ and that of spin glass $S_0^\mathrm{SG}$,

\begin{align}
    S_0^\mathrm{tot}(p) = f(p)S_0^\mathrm{SJ}(p) + [1-f(p)]S_0^\mathrm{SG}(p).
    \label{S_0}
\end{align}
This equation together with Eq. \ref{delta_S}, however, cannot give us a set of unique solutions, because $S_0^\mathrm{SJ}(p)$ and $S_0^\mathrm{SG}(p)$ can, in general, vary with $p$, and the analysis of $\Delta S$(0.5 K, 50 K) as a function of $p$ [shown in Fig. \ref{fig4}(b)] to estimate $S_0^\mathrm{SJ}(p)$ and $S_0^\mathrm{SG}(p)$ becomes an underconstrained problem.

To overcome this problem, we performed the entropy analysis by imposing two assumptions. The first assumption is based on the $p$-dependence of the correlation length, $\xi(p)$, reported by a previous neutron scattering study of SCGO \cite{Yang2015} in which $\xi(p)$ remains constant for $1.0 > p > 0.8$, i.e., it is robust against small nonmagnetic doping. At the same time, it linearly and gradually decreases with further decreasing $p$ below $p=0.8$. Since $\xi(p)$ is directly proportional to the spin-jam domain size, we assume that the spin-jam domain size is constant for $1.0 > p > 0.8$, and gradually changes in the same way as $\xi(p)$ does with decreasing $p$ for $p < 0.8$. The second assumption is that the zero-point entropy of spin jam scales with the perimeter of the spin jam domains as predicted by the spin-jam theory \cite{Klich2014}.

Now let us recall that our $C_\mathrm{mag}$ data yields the total zero-point entropy for $p = 0.98$ to be $S_0^\mathrm{tot} = 6.34$ Jmol$_\mathrm{Cr}^{-1}$K$^{-1}$. The zero-point entropy of the spin-glass state for $p = 0.98$ is likely to be very close to $S_\mathrm{max} = R\ln(2s+1) = 11.53$ Jmol$_\mathrm{Cr}^{-1}$K$^{-1}$ because, when the vacancy density is low, the spin glass is made of almost uncorrelated orphan spins that fluctuate nearly freely, resulting in $S_0^\mathrm{SG}(0.98) \approx S_\mathrm{max}$. Using Eq. \ref{S_0}, we obtained $S_0^\mathrm{SJ}(0.98) = 5.92$ Jmol$_\mathrm{Cr}^{-1}$K$^{-1}$, which is close to $S_0^\mathrm{tot}(0.98)$ as expected, as the magnetic glassy state for $p=0.98$ is predominantly a spin jam. For other $p$ values, we first estimated the perimeter of the spin jam domains with the experimental correlation length $\xi(p)$ of SCGO($p$) and scaled $S_0^\mathrm{SJ}(p)$ to $S_0^{\mathrm{SJ}}(p=0.98)$ according to the change in the perimeter of the magnetic domain (see Section IIA in \cite{SM} for details). Once $S_0^\mathrm{SJ}(p)$ is obtained, $S_0^\mathrm{SG}(p)$ for each $p$ was calculated using Eq. \ref{S_0}. Figure \ref{fig4}(c) shows the resulting $S_0^\mathrm{SJ}(p)$ and $S_0^\mathrm{SG}(p)$ for all $p > p_\mathrm{c}$. It is interesting to note that $S_0^\mathrm{SJ}(p)$ and $S_0^\mathrm{SG}(p)$ exhibit strikingly different behaviors with $p$. For $1.0 > p > 0.8$, the spin-jam zero-point entropy $S_0^\mathrm{SJ}(p)$ is much lower than $S_\mathrm{max} = 11.53$ Jmol$_\mathrm{Cr}^{-1}$K$^{-1}$. For $p < 0.8$, as $p$ decreases, on the other hand, $S_0^\mathrm{SJ}(p)$ gradually increases up to $0.81S_\mathrm{max}$ for $p = 0.51$, which is expected since the interaction-driven magnetic constraints become weaker as the domain size decreases. In contrast, as $p$ decreases after $p \sim 0.8$, $S_0^\mathrm{SG}(p)$ rapidly decreases down to $0.25S_\mathrm{max}$ for $p = 0.51$. The rapid decrease of $S_0^\mathrm{SG}(p)$ suggests that as the vacancy density in the magnetic lattice increases, the orphan spins begin to correlate with one another resulting in a smaller degeneracy as expected for canonical spin glasses \cite{Quilliam2007}. The value of $p \sim 0.8$, below which the orphan spins become strongly correlated, is also consistent with the neutron experimental results of SCGO \cite{Yang2015}.

To reaffirm the validity of our analysis, we estimated the typical perimeter of the spin-jam domain from the obtained value of $S_0^\mathrm{SJ} (0.98)$ based on the spin-jam theory \cite{Klich2014}. For $S_0^\mathrm{SJ}(0.98) \approx 5.9$ Jmol$_\mathrm{Cr}^{-1}$K$^{-1}$, the estimated number of bipyramids, denoted as $N_\mathrm{p}$, on the domain perimeter is approximately 1.5(1) (see Section IIB in \cite{SM} for detailed calculations). To put this value into context, we compared it with the magnetic domain size derived from the correlation length $\xi$ of a related system. In a single-crystal sample of SCGO(0.67), the measured $\xi$ using neutron scattering is  $4.6(2) \mathrm{\AA}$ \cite{Iida2012}. By considering the $p$ dependence of $\xi$ \cite{Yang2015}, we estimated $\xi$ to be 5.5(4) $\mathrm{\AA}$ for $p \sim 1$. This value closely matches the distance between the two centers of the nearest neighboring bipyramids of 5.85~\AA. $\xi = 5.5(4)~\mathrm{\AA}$ aligns with $N_\mathrm{p} \approx 1.5$ estimated from the spin jam theory, considering that $\xi$ is defined by the distance at which the spin correlation reduces to $e^{-1}$, while the spin jam theory using the transfer matrix formalism \cite{Klich2014} assumes the spin correlation being 1 within the magnetic domain.\\[3mm]

\section{Conclusion}\label{sec4}

In summary, our thermodynamic studies reveal that the low-temperature glassy state of BSZCGO is a mixture of the spin-jam and spin-glass states, characterized by the Halperin-Saslow modes and the localized two-level systems, respectively. The population ratio of the spin-glass state to the spin-jam state increases as $p$ decreases down to the percolation threshold, $p_\mathrm{c} = 0.5$. Furthermore, by quantitatively analyzing the magnetic zero-point entropy, we found that, as $p$ decreases, the zero-point entropy of the spin jam $S_0^\mathrm{SJ}(p)$ gradually increases, whereas that of the spin glass  $S_0^\mathrm{SG}(p)$ rapidly decreases below $p \sim 0.8$.  This work elucidates the coexistence of the two glassy states in the frustrated quantum magnetism.

\begin{acknowledgments}
J.Y. and S.-H.L. thank Dr. Matthias Thede and Dr. Andrey Zheludev for their help during some of our DC susceptibility measurements performed at Eidgenössische Technische Hochschule (ETH) Zurich. C.P. thanks Prof. Satoshi Kameoka for access to his X-ray diffractometer at Tohoku University. C.P. was supported by the DPST scholarship from the Institute for the Promotion of Teaching Science and Technology. Work at Mahidol University was supported in part by the National Research Council of Thailand Grant N41A640158 and the Thailand Center of Excellence in Physics. A.T. and S.-H.L. were supported by the US Department of Energy, Office of Science, Office of Basic Energy Sciences Award DE-SC0016144. W.-T.C. thanks the support by NSTC-Taiwan with project number 108-2112-M-002-025-MY3, TCECM project number 110-2124-M-002-019, and Academia Sinica iMATE grant number AS-iMATE-111-12. T.J.S. was supported by Grants-in-Aids for Scientific Research (JP22H00101, 19KK0069, 19H01834, 19K21839, 19H05824) from MEXT of Japan.
\end{acknowledgments}


%

\clearpage
\newpage
\widetext
\begin{center}
\textbf{\large Supplemental Material\\[3mm]Zero-point entropies of spin-jam and spin-glass states in a frustrated magnet}
\end{center}
\setcounter{equation}{0}
\setcounter{figure}{0}
\setcounter{table}{0}
\setcounter{section}{0}
\setcounter{page}{1}
\makeatletter
\renewcommand{\theequation}{S\arabic{equation}}
\renewcommand{\thefigure}{S\arabic{figure}}
\renewcommand{\thetable}{S\arabic{table}}
\renewcommand{\bibnumfmt}[1]{[S#1]}
\renewcommand{\citenumfont}[1]{S#1}

\section{Temperature-dependent molar heat capacity}

\begin{figure}[ht!]
    \centering
    \includegraphics[width=\textwidth]{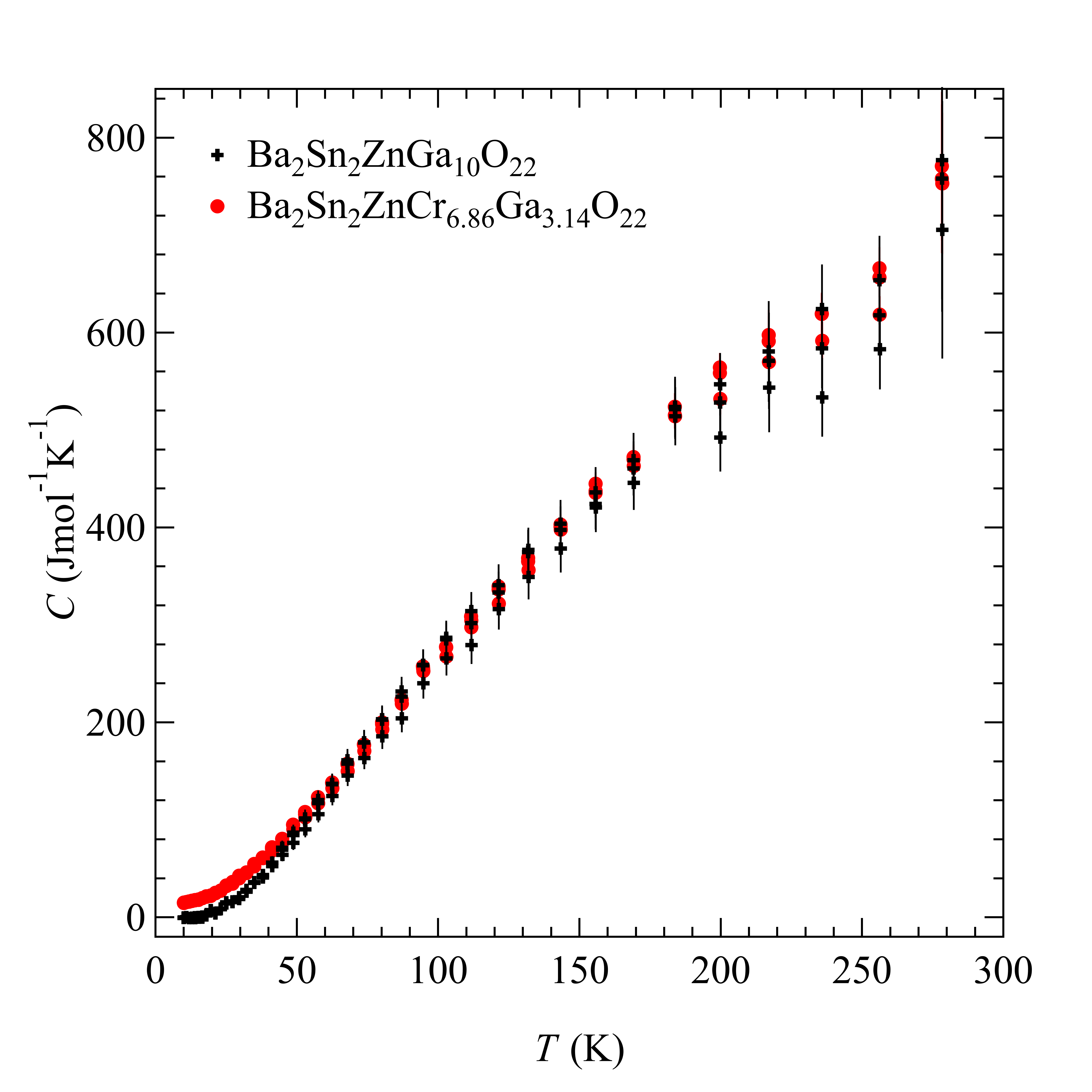}
    \caption{Temperature dependence of the total heat capacity $C(T)$ for $p=0.98$ and $p=0$ samples up to room temperature. Error bars represent one standard deviation.}
    \label{fig_S1-1}
\end{figure}

\begin{figure}[ht!]
    \centering
    \includegraphics[width=0.7\textwidth]{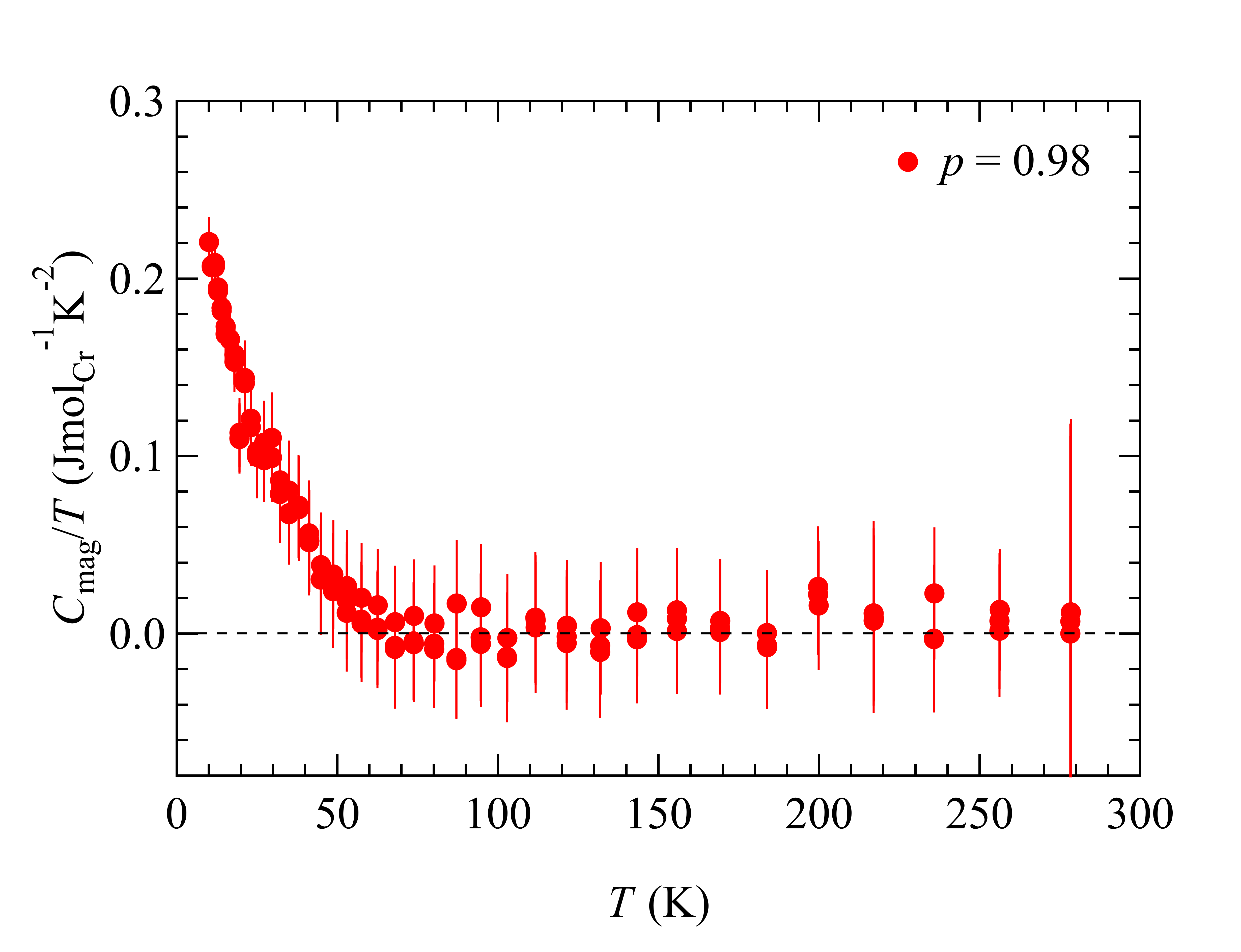}
    \caption{Temperature dependence of $C_\mathrm{mag}/T$ for the $p=0.98$ sample obtained from the results in Fig. \ref{fig_S1-1}. Error bars represent one standard deviation.}
    \label{fig_S1-2}
\end{figure}

Figure \ref{fig_S1-1} shows the temperature dependence of the total heat capacity $C(T)$ measured up to room temperature for $p=0$ and $p=0.98$ samples. From the figure, above $\sim$50 K, the heat capacity of both samples coincides with one another indicative of vanishing magnetic contribution to the magnetic sample ($p=0.98$) above such temperature. Figure \ref{fig_S1-2} depicts the temperature dependence of $C_\mathrm{mag}/T$, obtained from subtracting the phonon contribution measured on the non-magnetic sample ($p=0$) from the magnetic $p=0.98$ sample in Fig. \ref{fig_S1-1}, showing that $C_\mathrm{mag}/T$ of the $p=0.98$ sample tends to zero above $\sim$50 K. Hence, we conclude that there is no further significant increase in the magnetic entropy up to the room temperature. 

\begin{figure}[ht!]
    \centering
    \includegraphics[width=\textwidth]{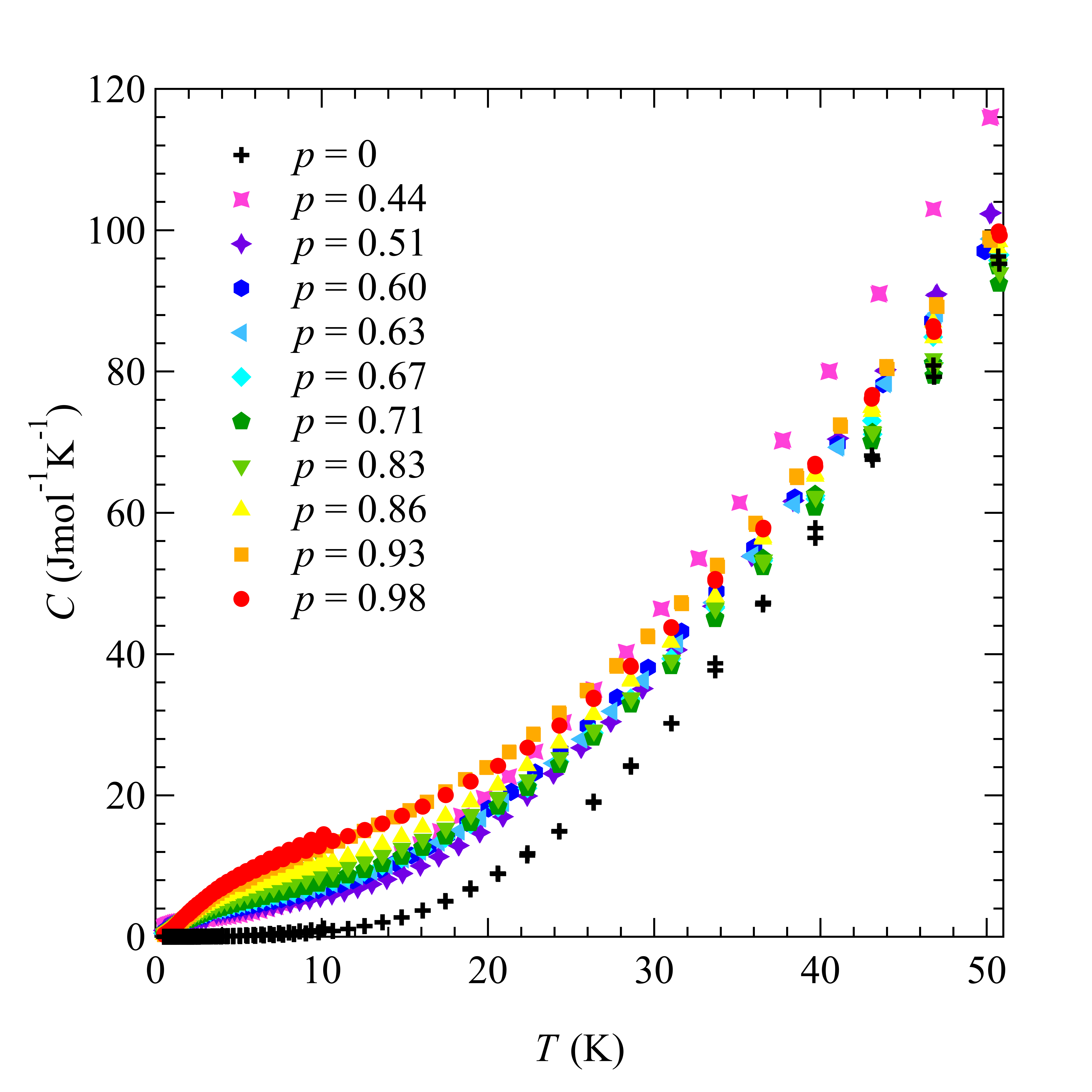}
    \caption{Temperature dependence of the total heat capacity $C(T)$ for all samples up to 50 K.}
    \label{fig_S1-3}
\end{figure}

Figure \ref{fig_S1-3} shows $C(T)$ of all samples measured from 0.5 K to 50 K. To obtain magnetic heat capacity, the magnetic $p\neq0$ sample's heat capacity is subtracted by the heat capacity of the non-magnetic $p=0$ sample, where the non-magnetic data are interpolated before the subtraction. After the subtraction, the data in the overlapping region taken with He-3 and He-4 options are averaged, weighted by errors as shown in Fig. 3(a).

\section{Calculation of spin-jam zero-point entropy, $S_0^\mathrm{SJ}$}

\subsection{The $p$ dependence of $S_0^\mathrm{SJ}$}

The calculation of the $p$-dependent zero-point entropy of spin jam $S_0^\mathrm{SJ}(p)$ is performed based on the spin jam theory of triangular network of bipyramids and the previously measured spin correlation lengths of the related compound SrCr$_{9p}$Ga$_{12-9p}$O$_{19}$ (SCGO($p$)) with a similar structure as a function of $p$. Then, using Eq. (4) in the main text, the $p$ dependence of spin glass's zero-point entropy $S_0^\mathrm{SG}(p)$ is obtained from $S_0^\mathrm{SJ}(p)$ and $S_0^\mathrm{tot}(p)$.

The $p$-dependent normalized perimeter length is obtained from the $p$-dependent normalized correlation length measured from SCGO($p$). The correlation length is found to remain constant for $0.8 < p < 1.0$ and it linearly decreases with $p$ below $p \sim 0.8$. To calculate the normalized perimeter length for each domain size, we suppose that the correlation length corresponds to the radius of the domain. We start with a circle of radius 1 unit. Then, we fill this circle with smaller circles representing smaller domains in a close-packed pattern. The total perimeter length of all circles with the same radius within the largest circle is, then, calculated. As a result, we can determine the total perimeter length of domains with different sizes covering the same total area. Hence, the total perimeter length of a larger domain, normalized with the total area, is smaller. According to the theory, the ground-state configurational entropy of spin jam scales with the normalized perimeter of the domain. Thus, the $p$ dependence of the spin jam's zero-point entropy $S_0^\mathrm{SJ}(p)$ can be determined from that of the normalized domain perimeter.

\subsection{Perimeter scaling of $S_0^\mathrm{SJ}$}

$S_0^\mathrm{SJ}$ can be calculated using the formula $S^\mathrm{SJ}_0 = k\ln\Omega/n = R\ln\Omega/N$, where $\Omega$ is the number of all possible ground-state spin configurations, $n$ the number of moles of spins, and $N$ the number of spins. As detailed in Ref. [37], to obtain a ground-state spin configuration, a sign state must be imposed with a color configuration. As there are 6 different color configurations, the number of all spin configurations, $\Omega$, is hence equal to $6\cdot\Omega_\mathrm{sign}$, where $\Omega_\mathrm{sign}$ is the total number of sign states. $\Omega_\mathrm{sign}$ is numerically calculated in Ref. [37], where $\log_{10}\Omega_\mathrm{sign}$ increases linearly with the number of bipyramids positioned on the perimeter $N_\mathrm{p}$, i.e., $\log_{10}\Omega_\mathrm{sign} = mN_\mathrm{p} + c$, as shown in the inset of Fig.~2(b) in Ref. [37].

A linear fit to $\log_{10}\Omega_\mathrm{sign}$ vs. $N_\mathrm{p}$ yields $m \approx 0.26$ and $c \approx -0.78$ as shown in Fig. \ref{fig_S2-1}. Therefore, we obtain $\ln\Omega = \ln(6\cdot\Omega_\mathrm{sign}) = \ln(6\cdot10^{mN_\mathrm{p} + c}) \approx \alpha N_\mathrm{p}$ where $\alpha = m\ln 10 \approx 0.60$. With this result, we obtained $S^\mathrm{SJ}_0 \approx R(\alpha N_\mathrm{p})/N$. For a circular-shape domain, the number of bipyramids in the domain can be approximated by $N_\mathrm{p}^2/4\pi$. As a bipyramid contains 7 spins, the number of spins $N$ in the typical domain is, therefore, $N \approx 7N_\mathrm{p}^2/4\pi$, and hence $S^\mathrm{SJ}_0 = 4\pi \alpha R/7N_\mathrm{p}$. Using the experimental value of $S^\mathrm{SJ}_0 = 5.9(5)$ Jmol$_\mathrm{Cr}^{-1}$K$^{-1}$ for $p = 0.98$, we obtain $N_\mathrm{p} = 1.5(1)$ for the magnetic domain.

\section{Structural refinements}

The X-ray diffraction results are analyzed using Rietveld refinements. To demonstrate the capability of X-ray diffraction in determining the occupancy of Cr$^{3+}$ ions in a magnetic sample, we compare the diffraction peaks of three different samples within the range of $63.0^\circ < 2\theta < 64.5^\circ$, normalizing them to have the same highest peak count, as shown in Figure \ref{fig_S3-12}. In the figure, besides varying peak positions, each sample also exhibits distinct intensities within this range. These differences in peak positions are likely attributed to variations in the lattice parameters of each sample. Conversely, simulations based on similar lattice and atomic parameters but differing Cr$^{3+}$ occupancies, as illustrated in Figure \ref{fig_S3-13}, indicate that the disparities in intensities observed in Figure \ref{fig_S3-12} are a result of varying Cr$^{3+}$ occupancies.

The initial crystal-structure parameters used in the refinements are taken from Ref. [51]. Figures \ref{fig_S3-1}-\ref{fig_S3-11} show the results of Rietveld refinements for each sample along with corresponding optimal atomic parameters in Tables \ref{structure_0p00}-\ref{structure_0p98}. The refined lattice parameters, $a = b$ and $c$, are plotted against $p$ in Fig. \ref{fig_S3-14}. To minimize errors of the occupancies of the Cr$^{3+}$ ions in the magnetic samples, other atomic and lattice parameters were fixed in the final refinement. The $p$ value for each sample is obtained by averaging the occupancies of the two Cr$^{3+}$ sites weighted by the corresponding site multiplicity.

\newpage

\begin{figure}[ht!]
    \centering
    \includegraphics[width=\textwidth]{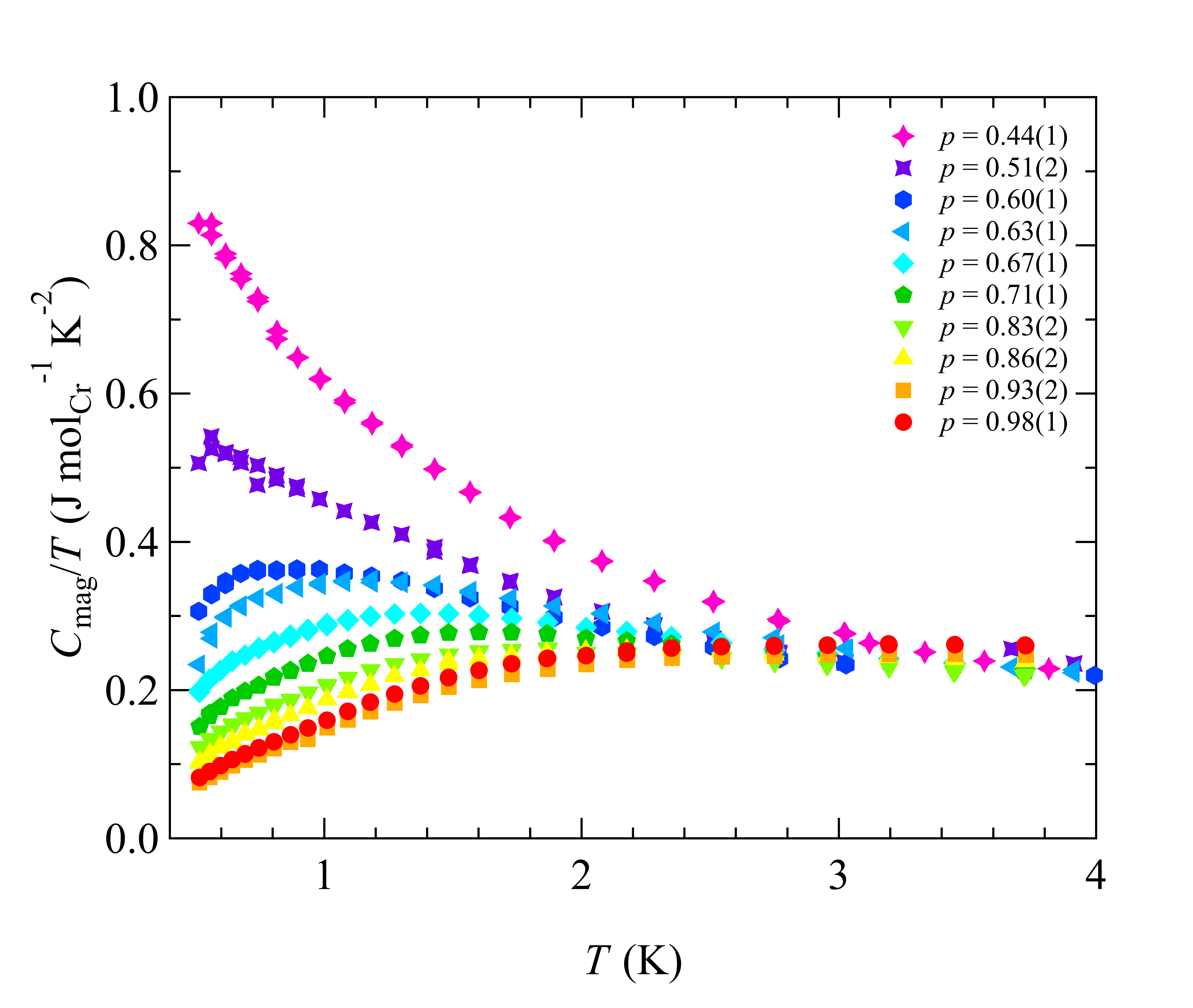}
    \caption{Low-temperature region of $C_\mathrm{mag}/T$ for all samples.}
    \label{fig_S1-4}
\end{figure}

\begin{figure}[ht!]
    \centering
    \includegraphics[width=\textwidth]{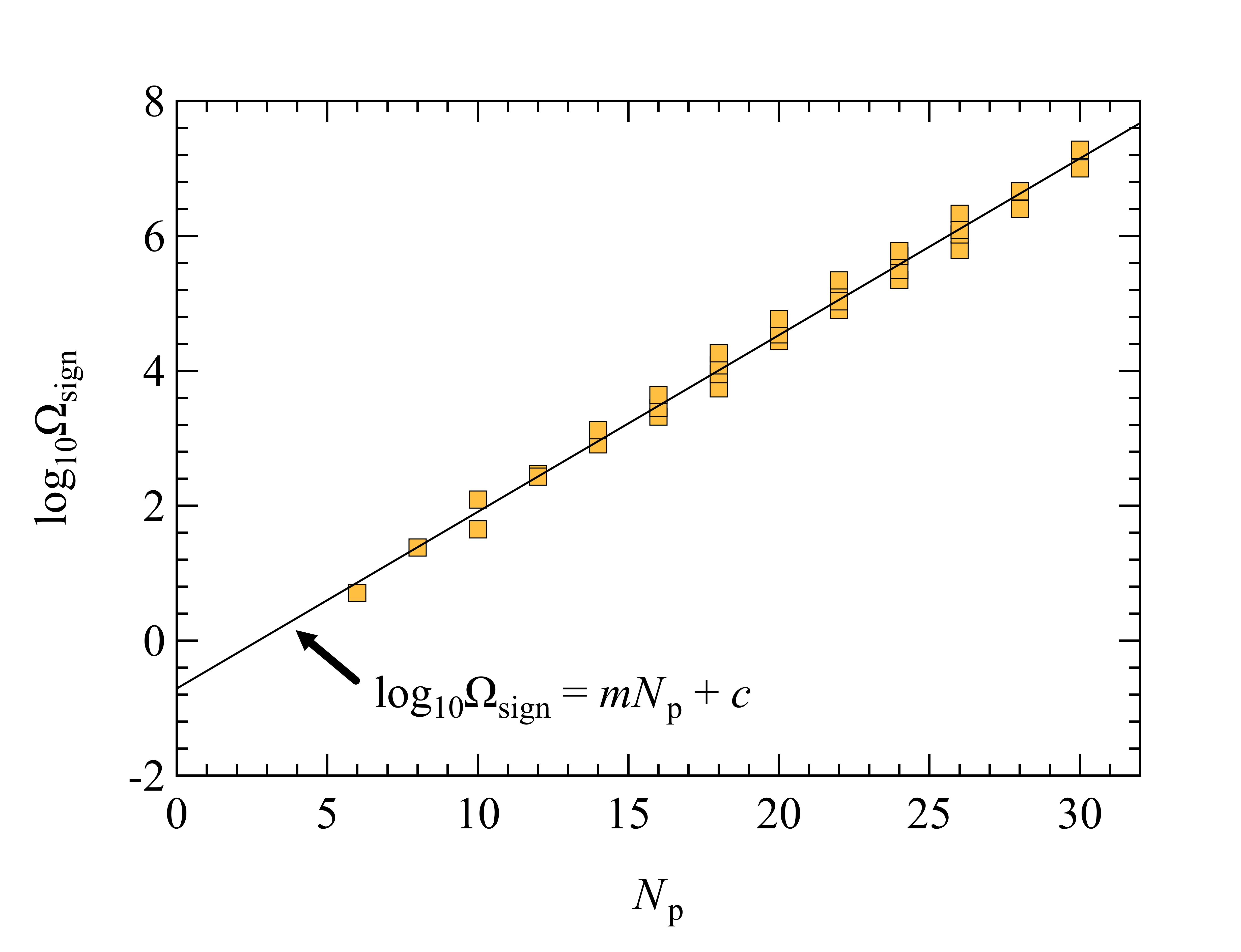}
    \caption{Scaling of $\log_{10}\Omega_\textrm{sign}$ with the domain perimeter $N_\textrm{p}$ in a unit of bipyramids. $\Omega_\textrm{sign}$ is the number of sign states as described in Ref. [37]. This plot is reproduced from the inset of Fig. 2(b) in Ref. [37].}
    \label{fig_S2-1}
\end{figure}

\begin{figure}[ht!]
    \centering
    \includegraphics[width=\textwidth]{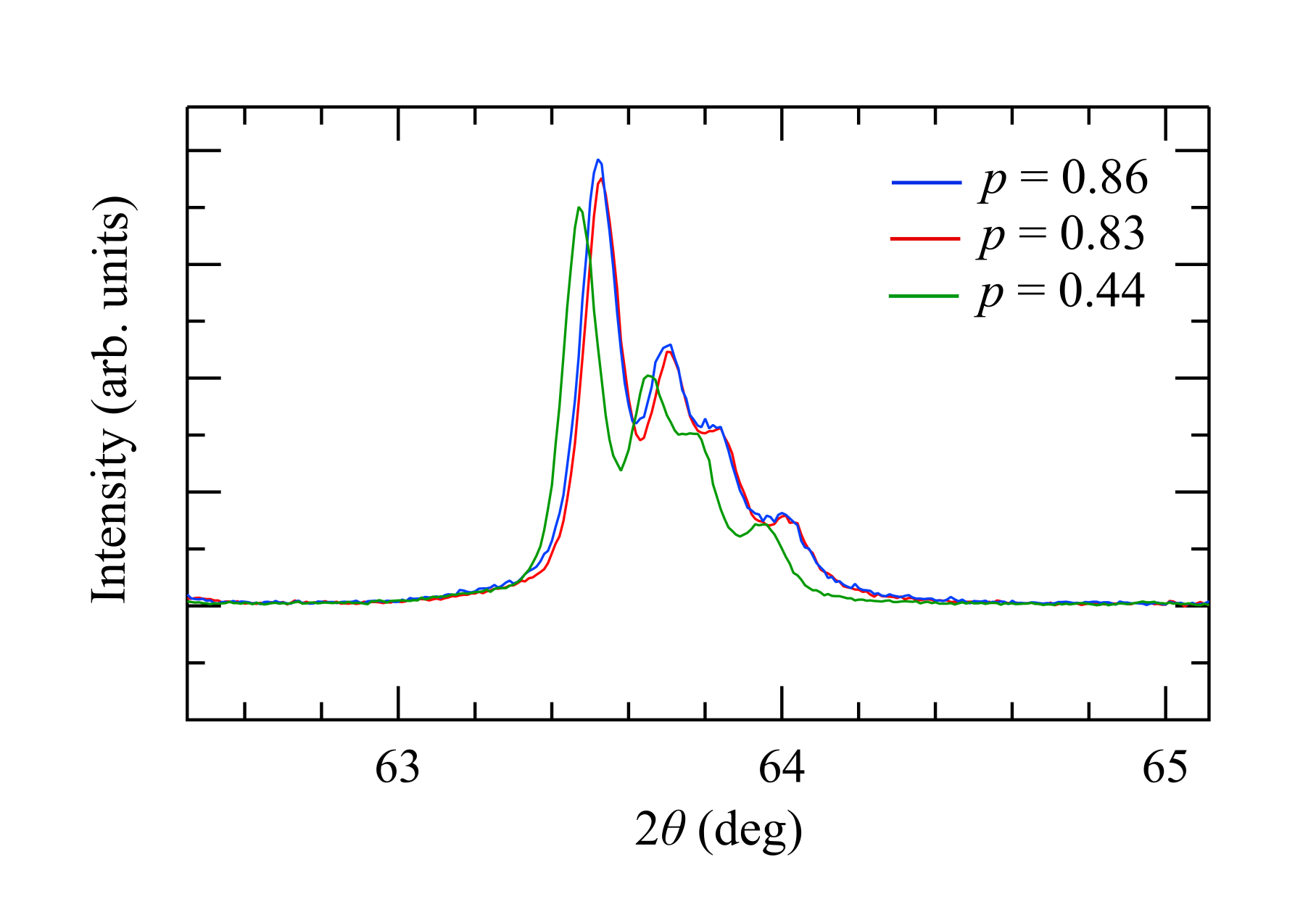}
    \caption{Comparison of diffraction peaks of 3 different samples in $63.0 ^\circ < 2\theta < 64.5 ^\circ$ region. The $p$ values are obtained from the Rietveld refinements.}
    \label{fig_S3-12}
\end{figure}

\begin{figure}[ht!]
    \centering
    \includegraphics[width=\textwidth]{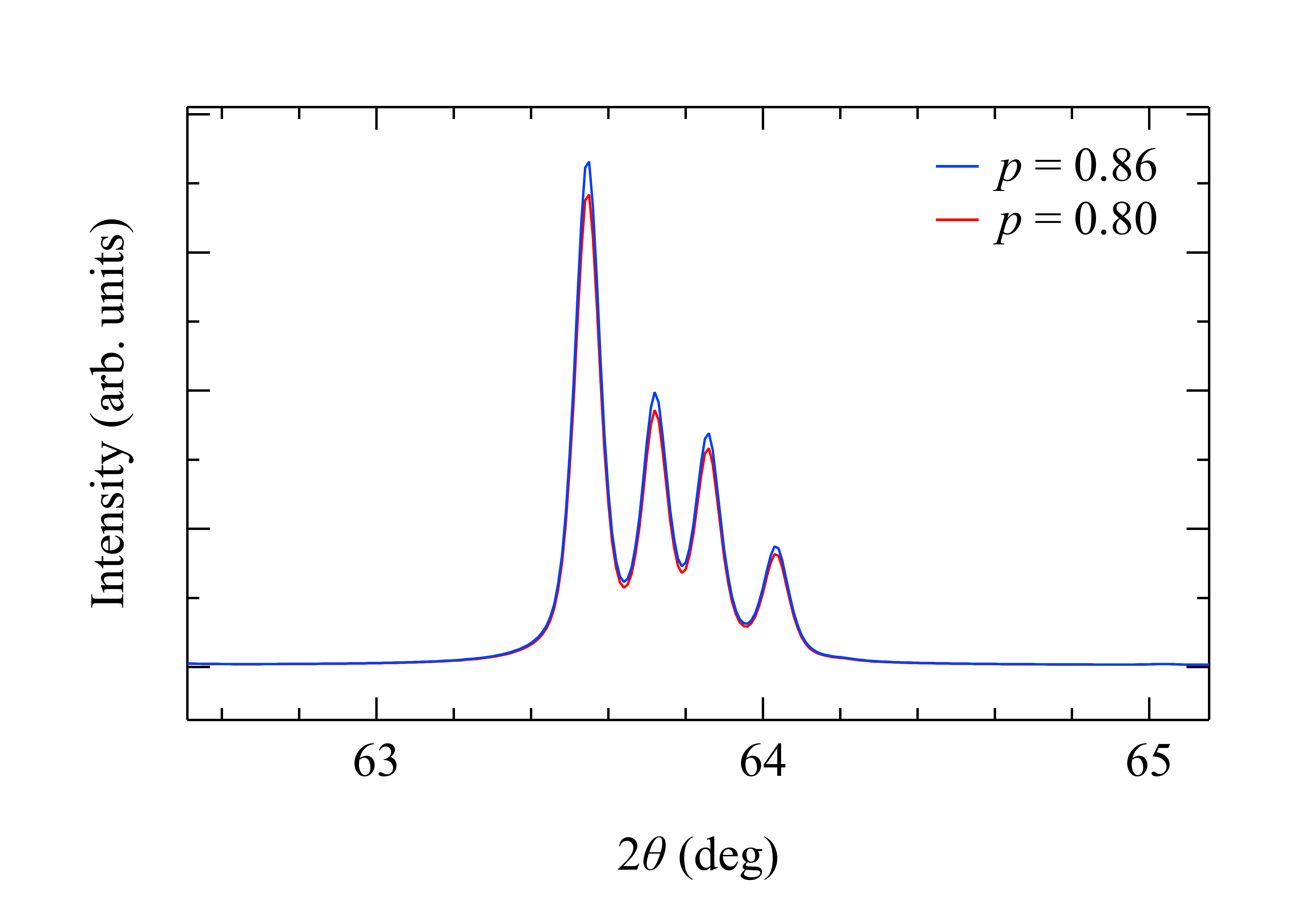}
    \caption{Comparison of simulated diffraction peaks in the same region as in Fig. \ref{fig_S3-12} based on two models with similar lattice and atomic parameters but with distinct Cr$^{3+}$ occupancies.}
    \label{fig_S3-13}
\end{figure}

\begin{figure}[ht!]
    \centering
    \includegraphics[width=\textwidth]{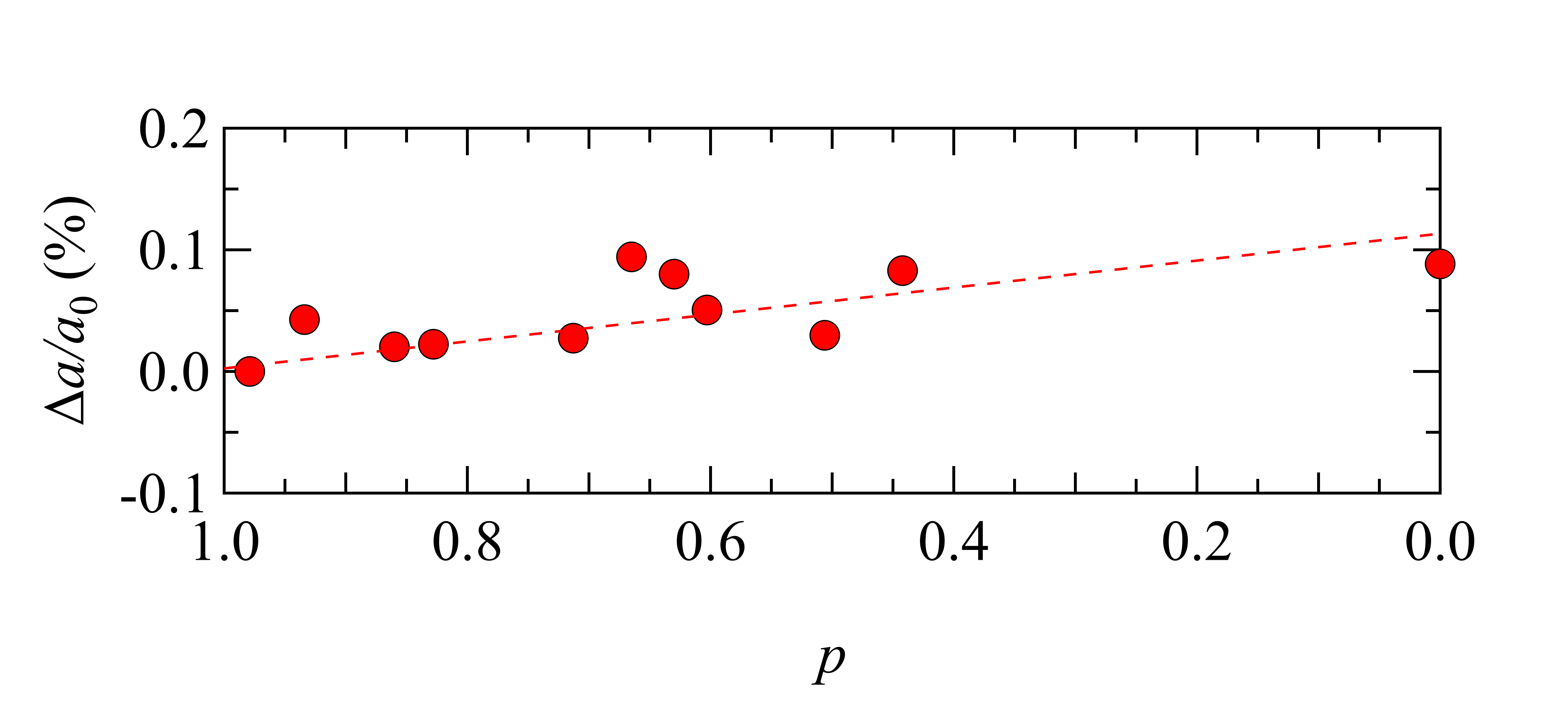}
    \includegraphics[width=\textwidth]{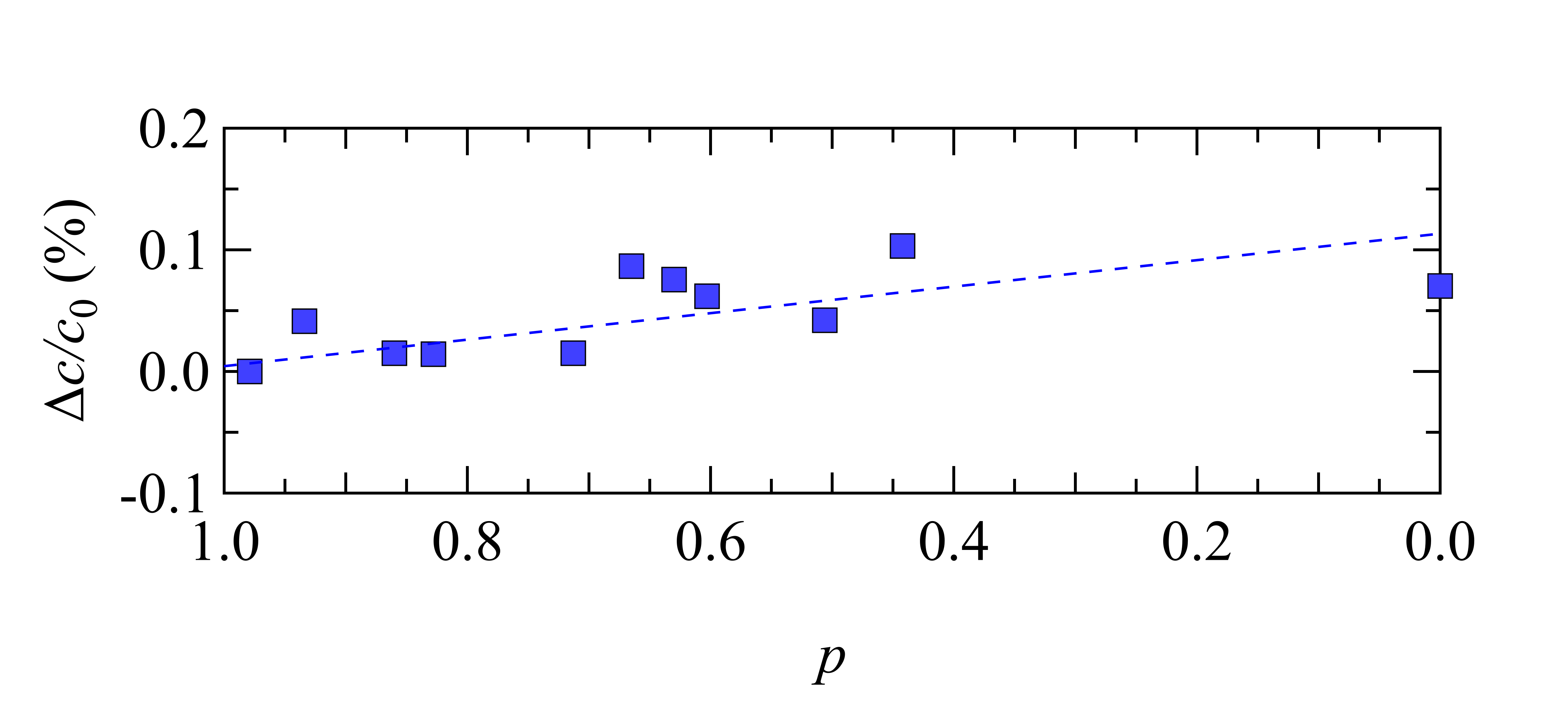}
    \caption{The lattice parameters $a$ ($a = b$) and $c$ obtained from the refinements. The results are shown as percentage change with respect to that of the $p=0.98$ sample, $a_0$ and $c_0$. Lines are guides to the eye.}
    \label{fig_S3-14}
\end{figure}

\begin{figure}[ht!]
    \centering
    \includegraphics[width=\textwidth]{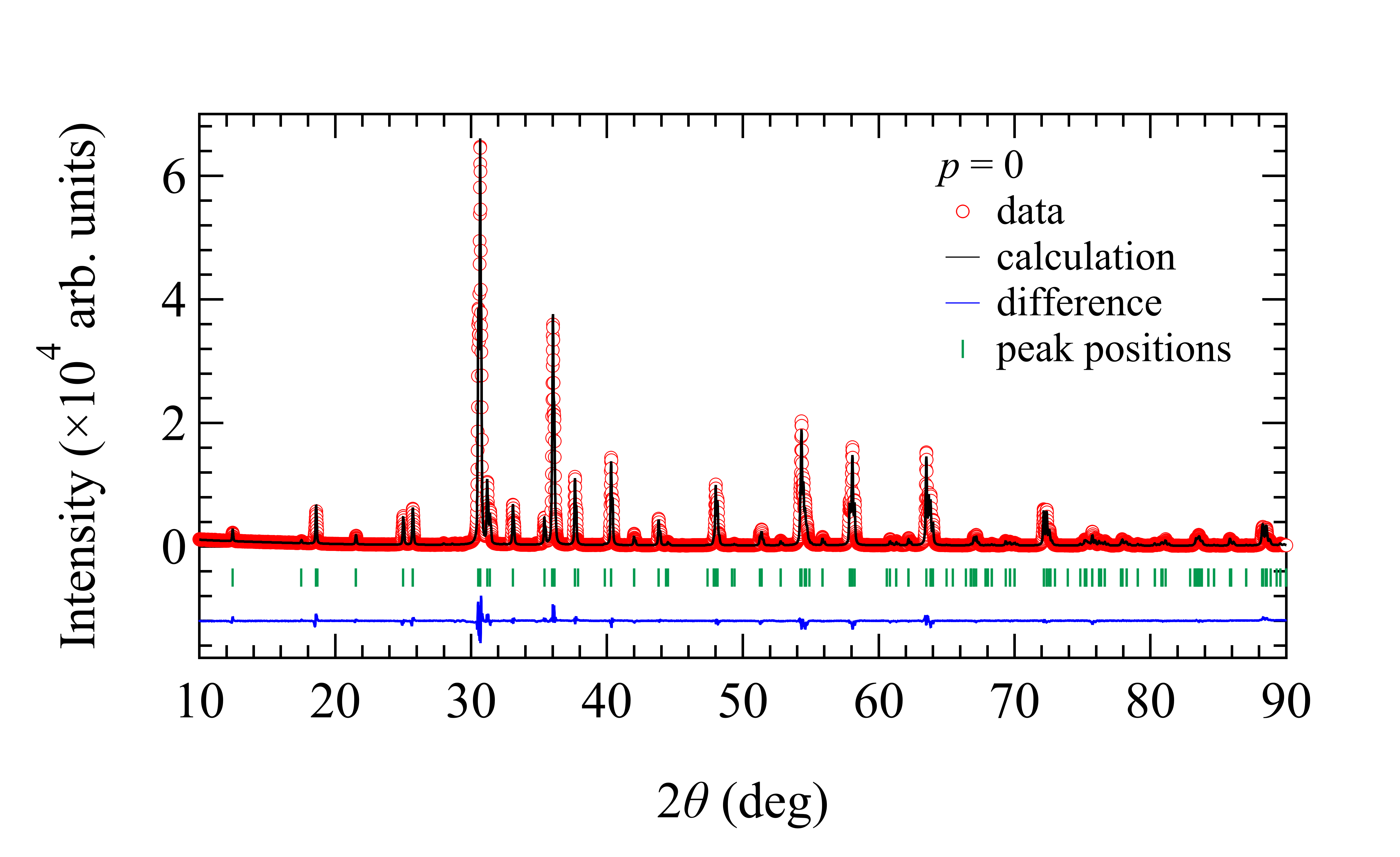}
    \caption{Refinement result of the $p=0$ sample.}
    \label{fig_S3-1}
\end{figure}

\begin{figure}[ht!]
    \centering
    \includegraphics[width=\textwidth]{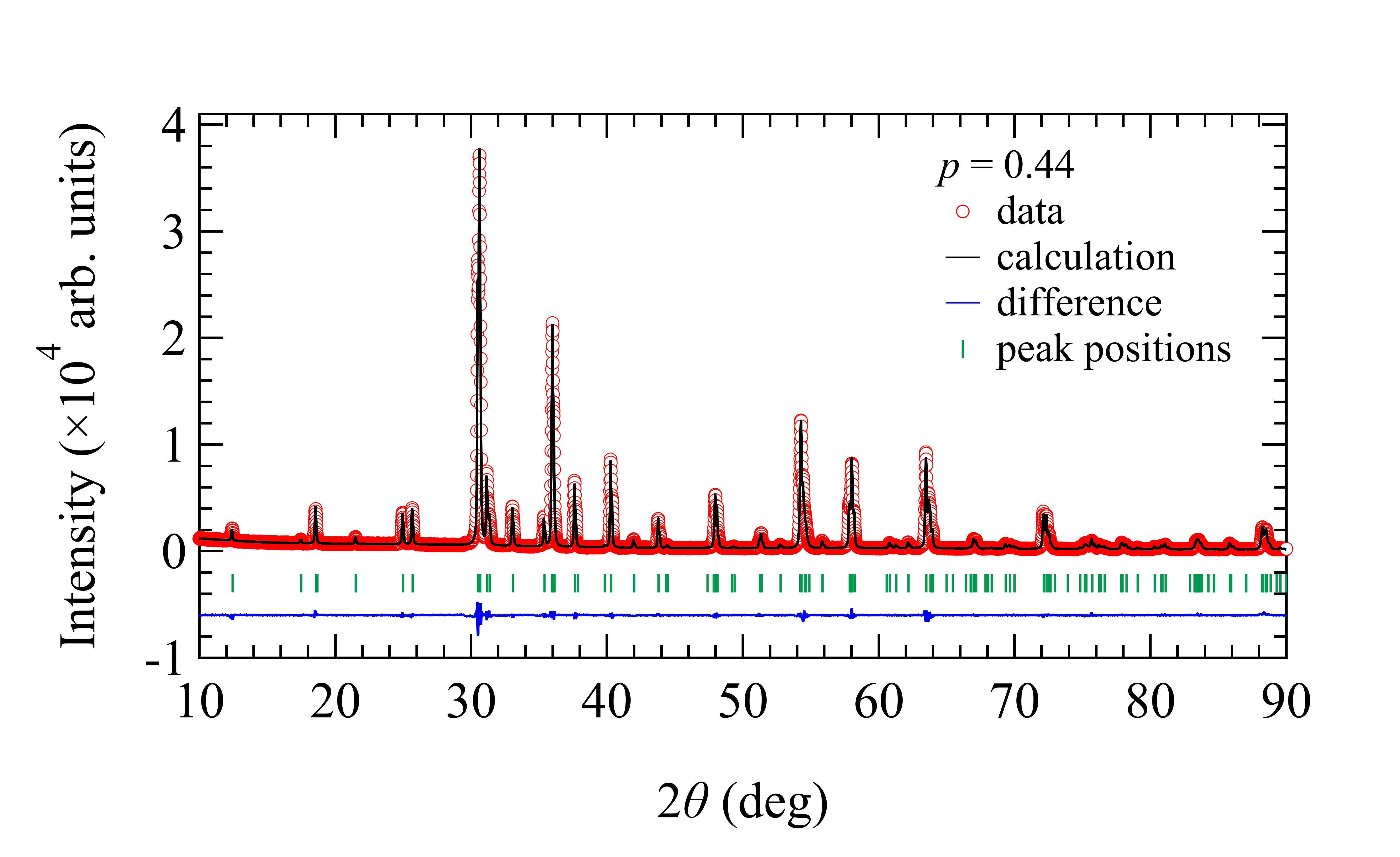}
    \caption{Refinement result of the $p=0.44$ sample.}
    \label{fig_S3-2}
\end{figure}

\begin{figure}[ht!]
    \centering
    \includegraphics[width=\textwidth]{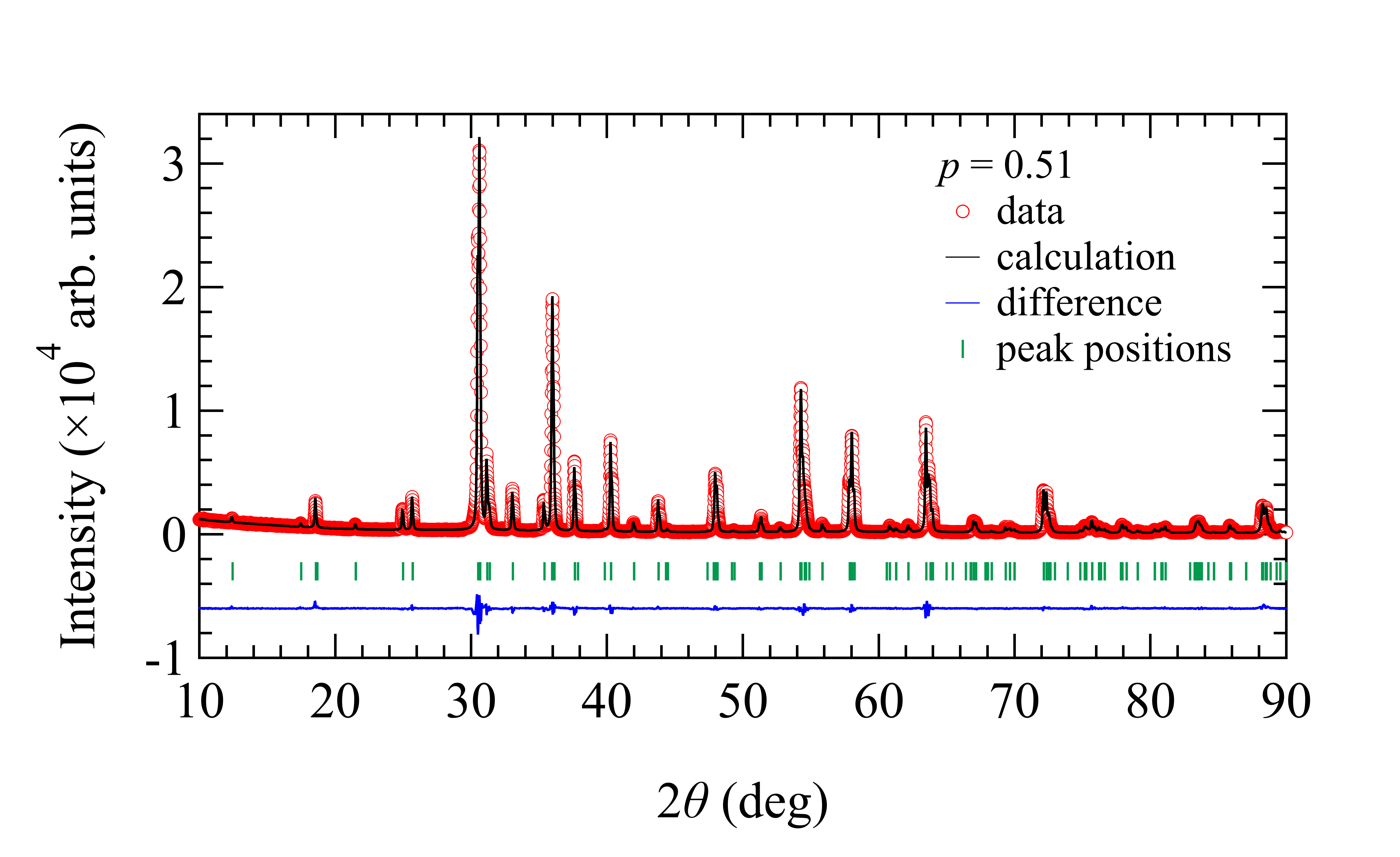}
    \caption{Refinement result of the $p=0.51$ sample.}
    \label{fig_S3-3}
\end{figure}

\begin{figure}[ht!]
    \centering
    \includegraphics[width=\textwidth]{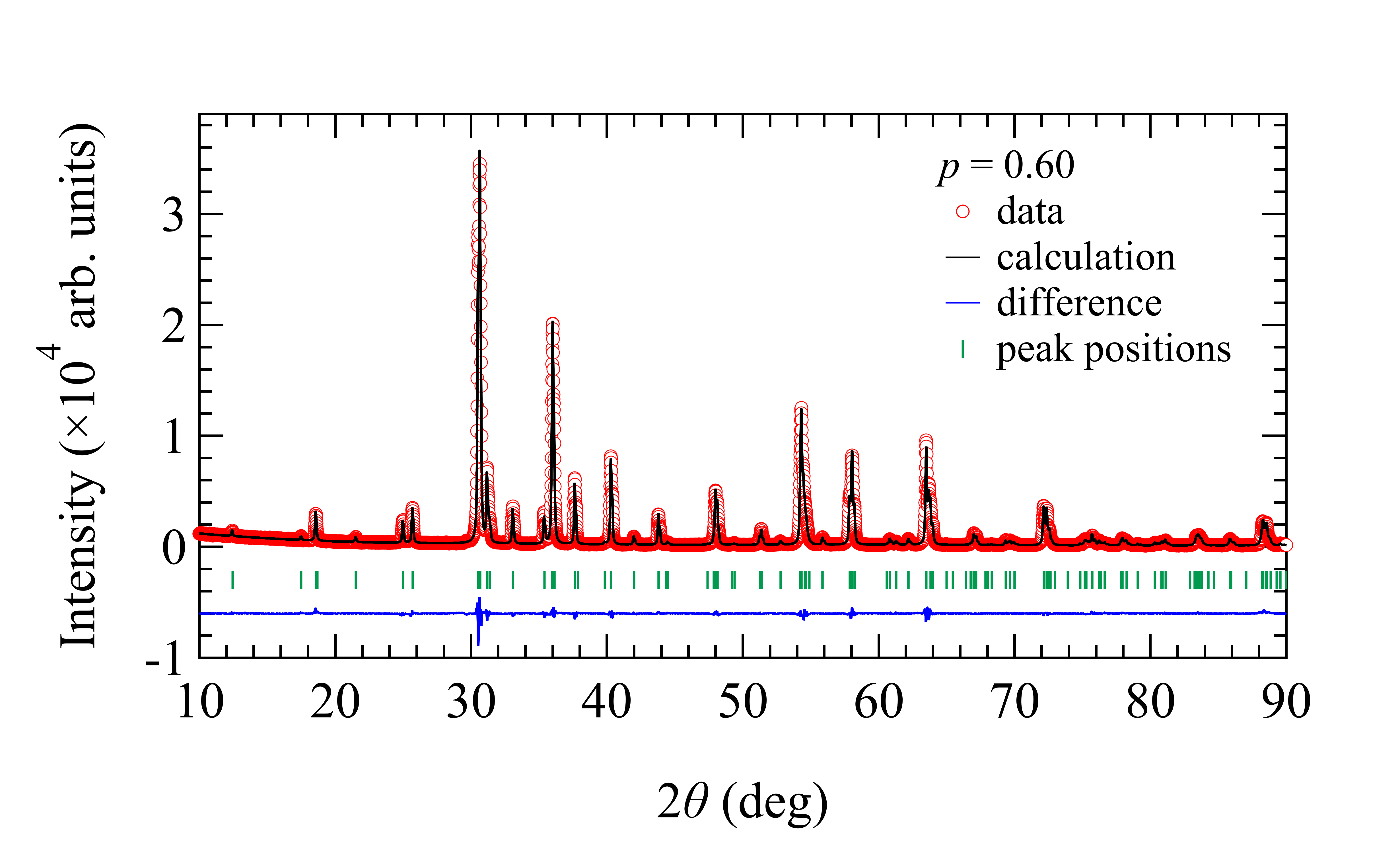}
    \caption{Refinement result of the $p=0.60$ sample.}
    \label{fig_S3-4}
\end{figure}

\begin{figure}[ht!]
    \centering
    \includegraphics[width=\textwidth]{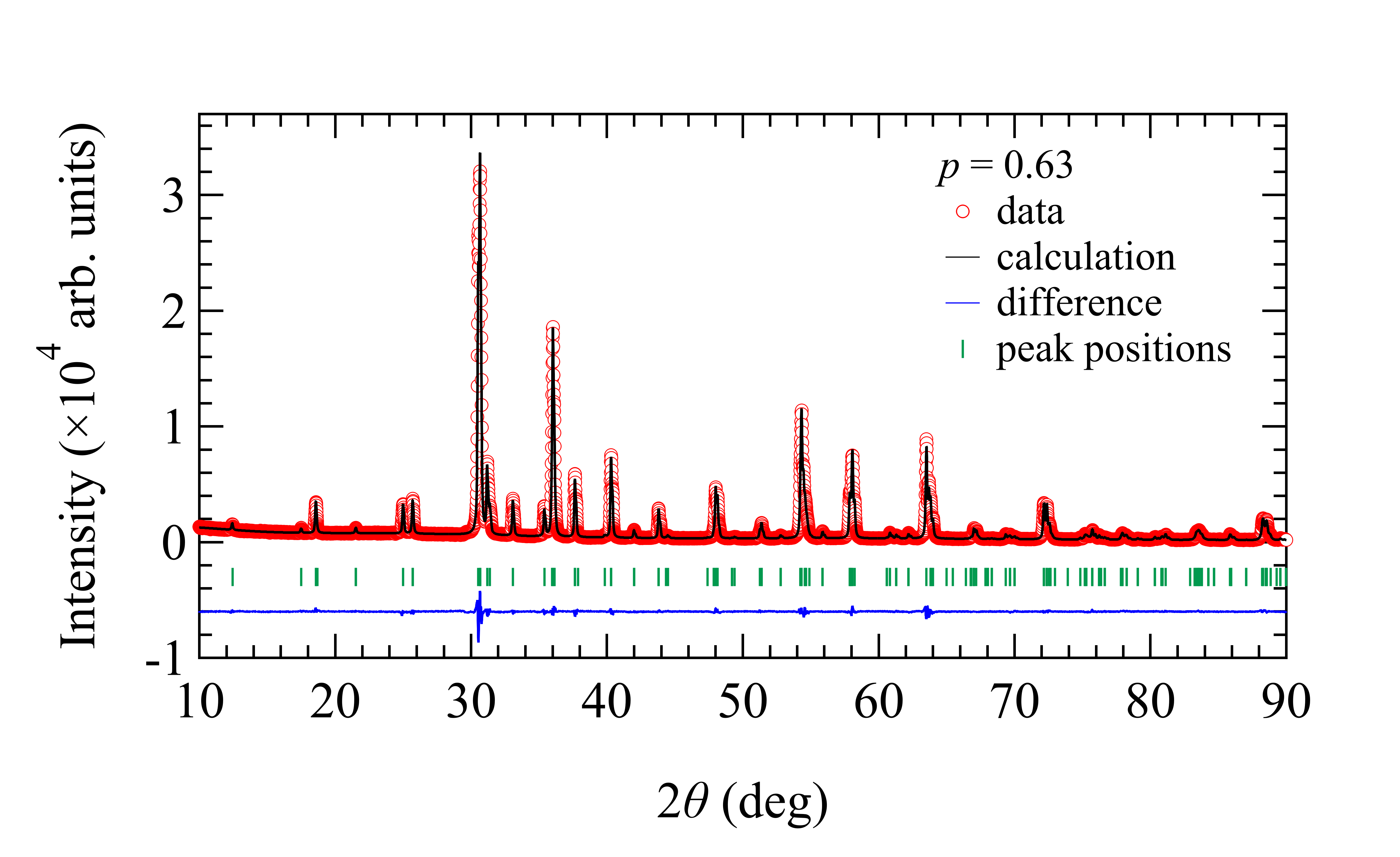}
    \caption{Refinement result of the $p=0.63$ sample.}
    \label{fig_S3-5}
\end{figure}

\begin{figure}[ht!]
    \centering
    \includegraphics[width=\textwidth]{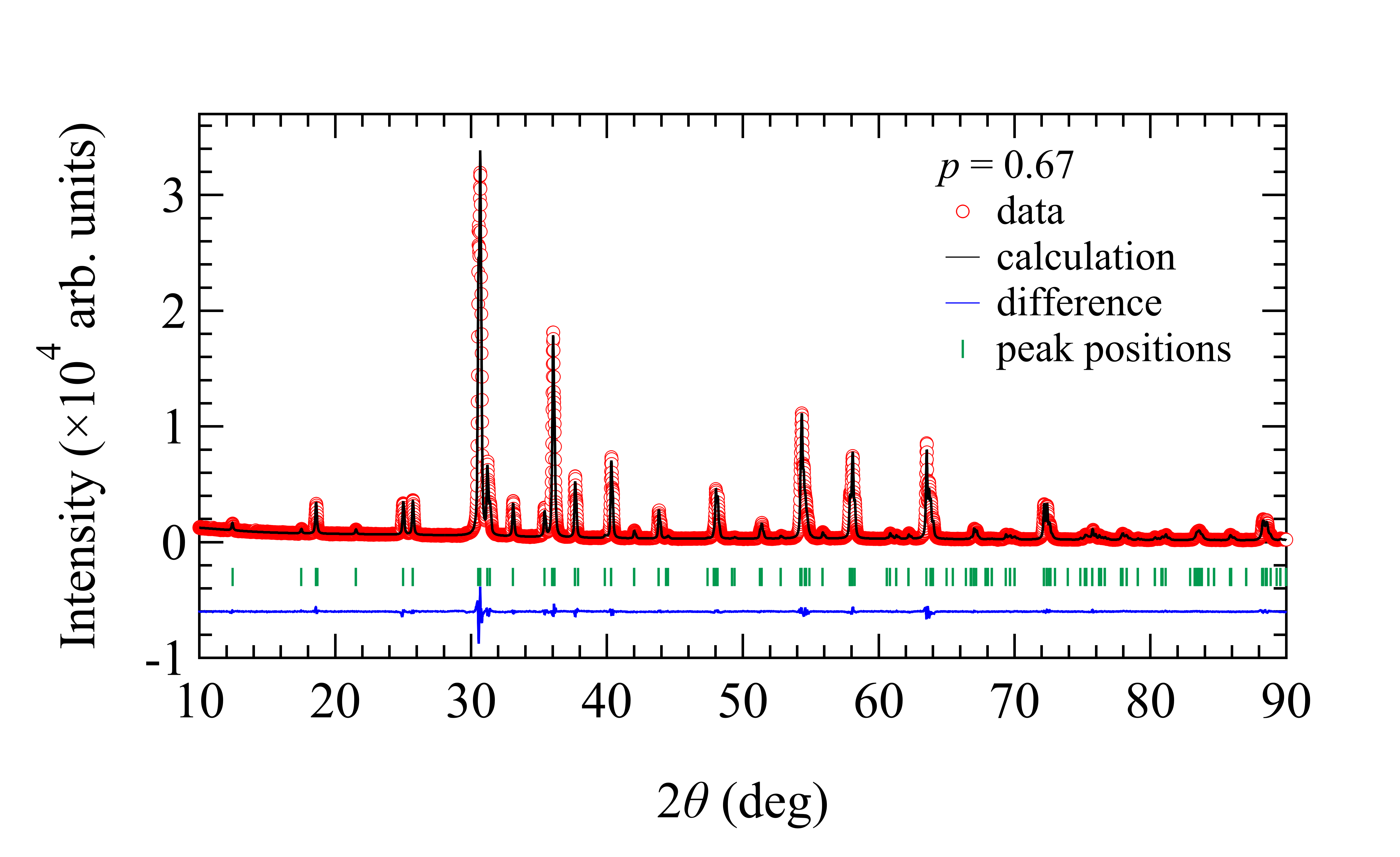}
    \caption{Refinement result of the $p=0.67$ sample.}
    \label{fig_S3-6}
\end{figure}

\begin{figure}[ht!]
    \centering
    \includegraphics[width=\textwidth]{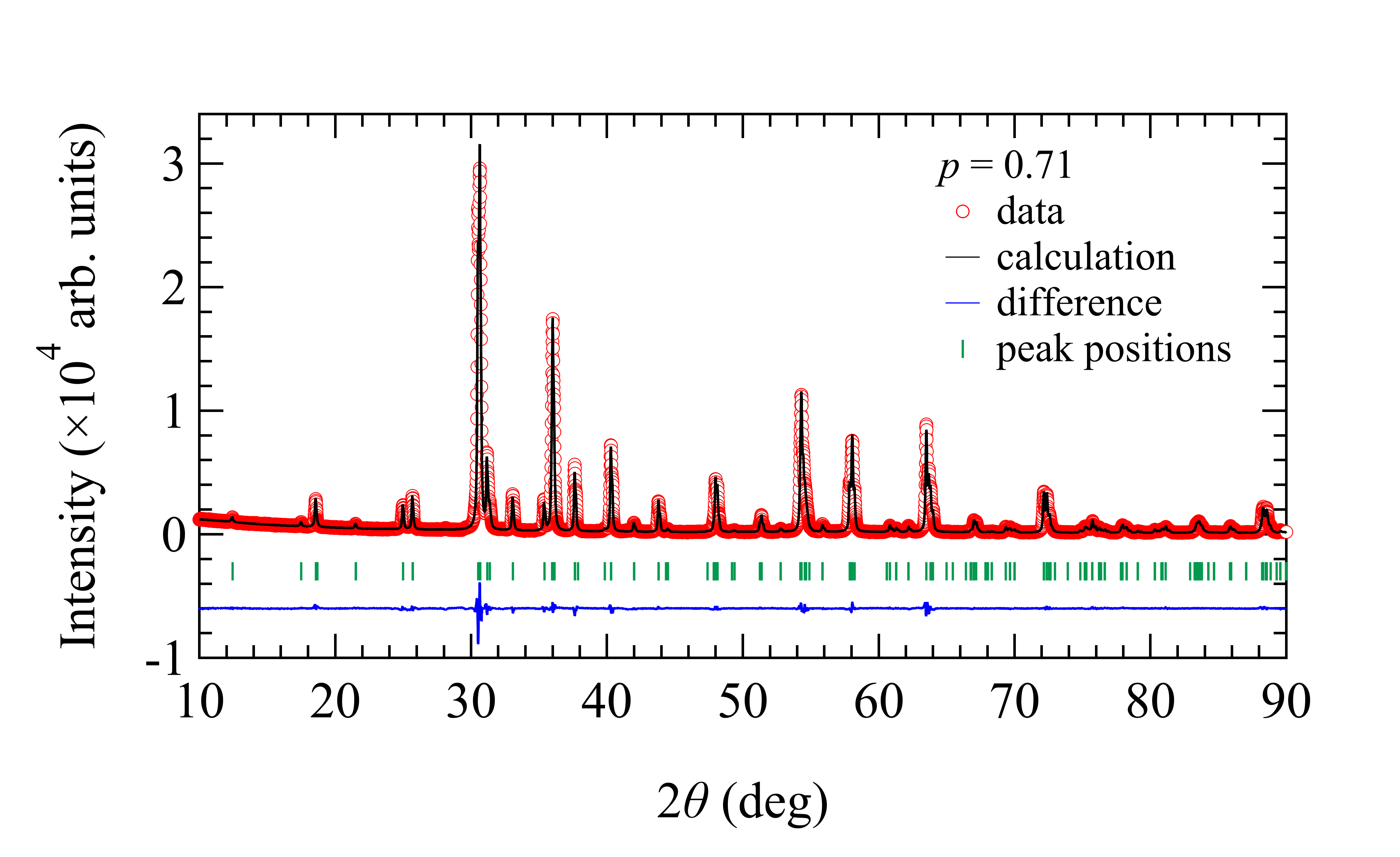}
    \caption{Refinement result of the $p=0.71$ sample.}
    \label{fig_S3-7}
\end{figure}

\begin{figure}[ht!]
    \centering
    \includegraphics[width=\textwidth]{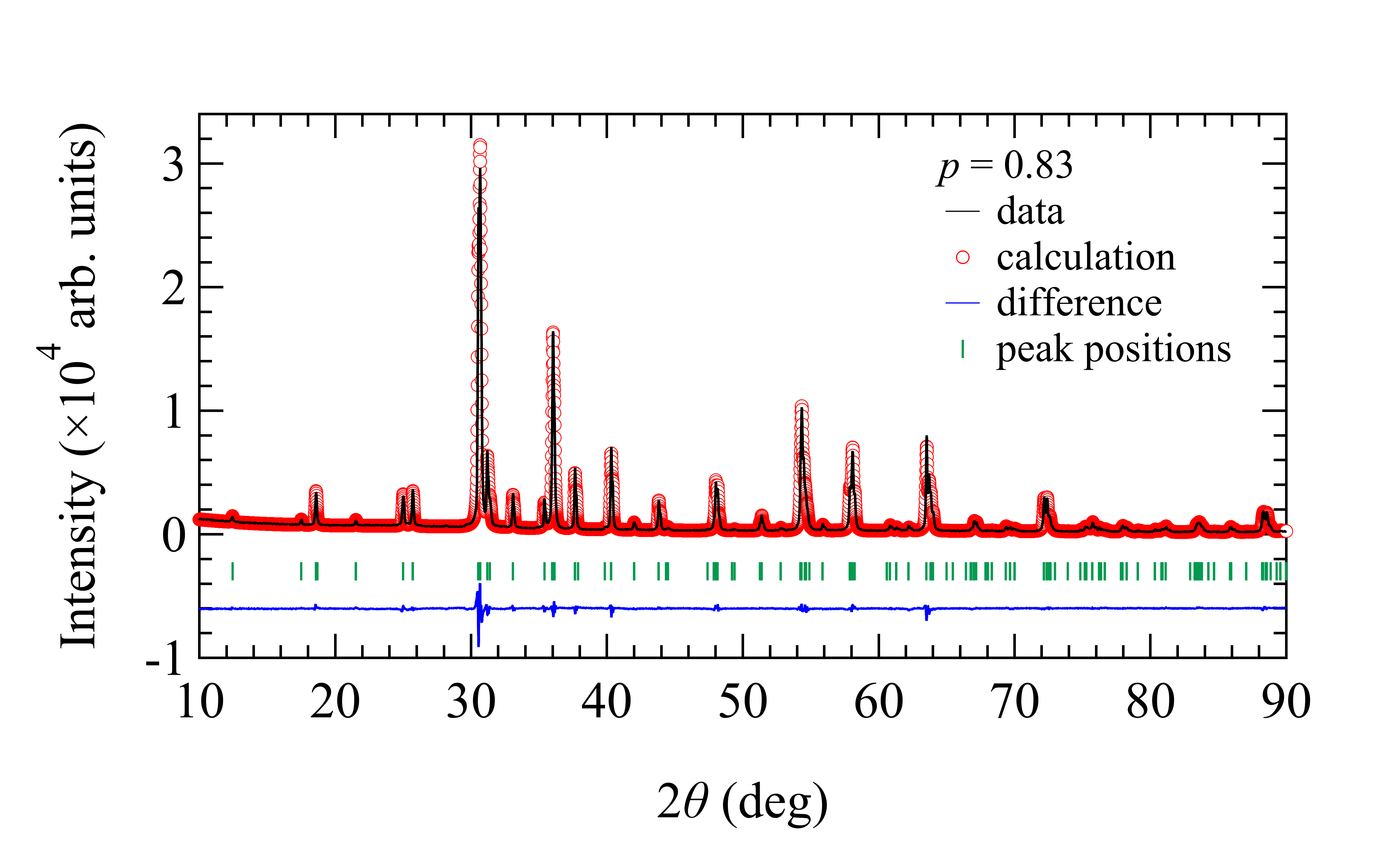}
    \caption{Refinement result of the $p=0.83$ sample.}
    \label{fig_S3-8}
\end{figure}

\begin{figure}[ht!]
    \centering
    \includegraphics[width=\textwidth]{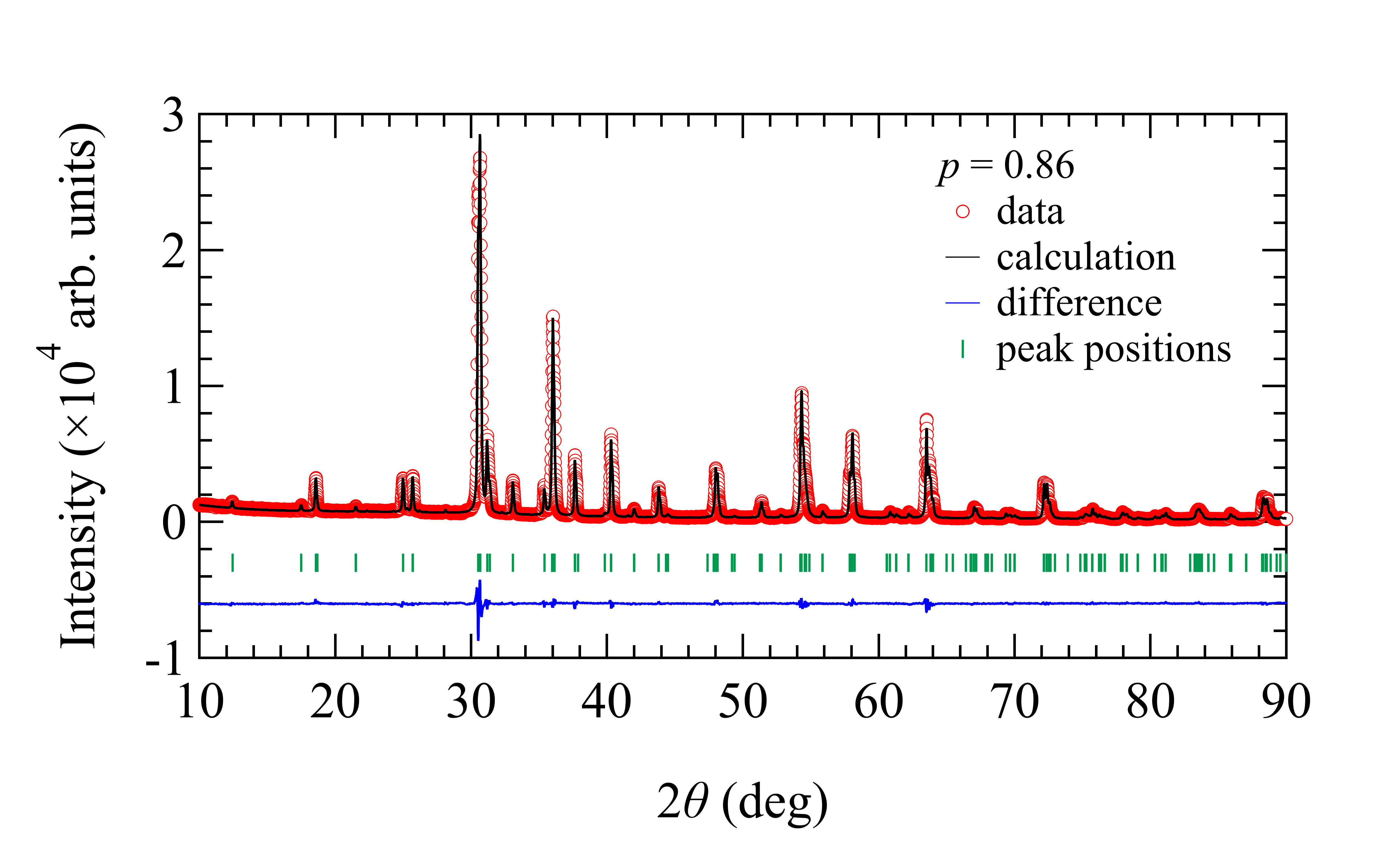}
    \caption{Refinement result of the $p=0.86$ sample.}
    \label{fig_S3-9}
\end{figure}

\begin{figure}[ht!]
    \centering
    \includegraphics[width=\textwidth]{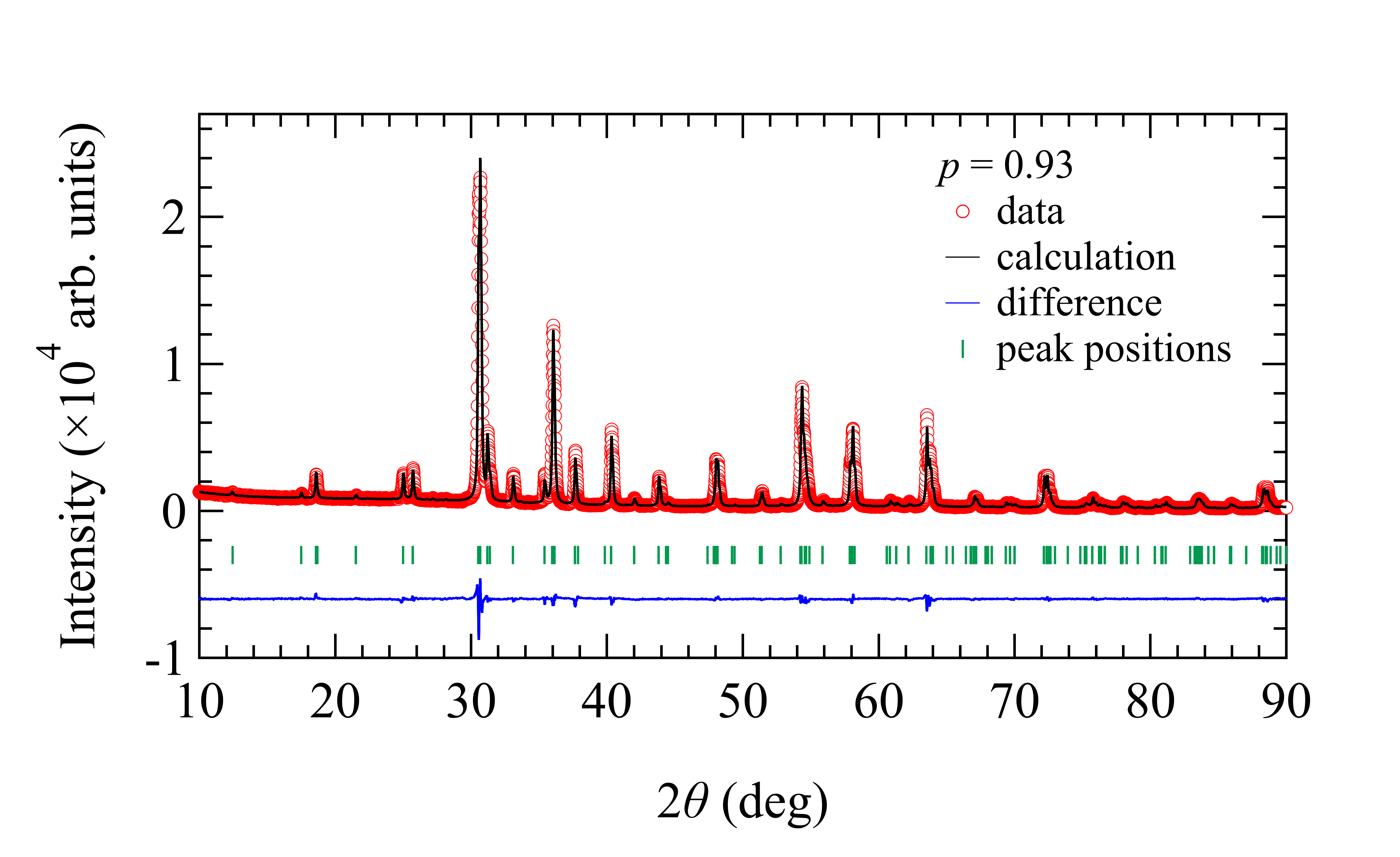}
    \caption{Refinement result of the $p=0.93$ sample.}
    \label{fig_S3-10}
\end{figure}

\begin{figure}[ht!]
    \centering
    \includegraphics[width=\textwidth]{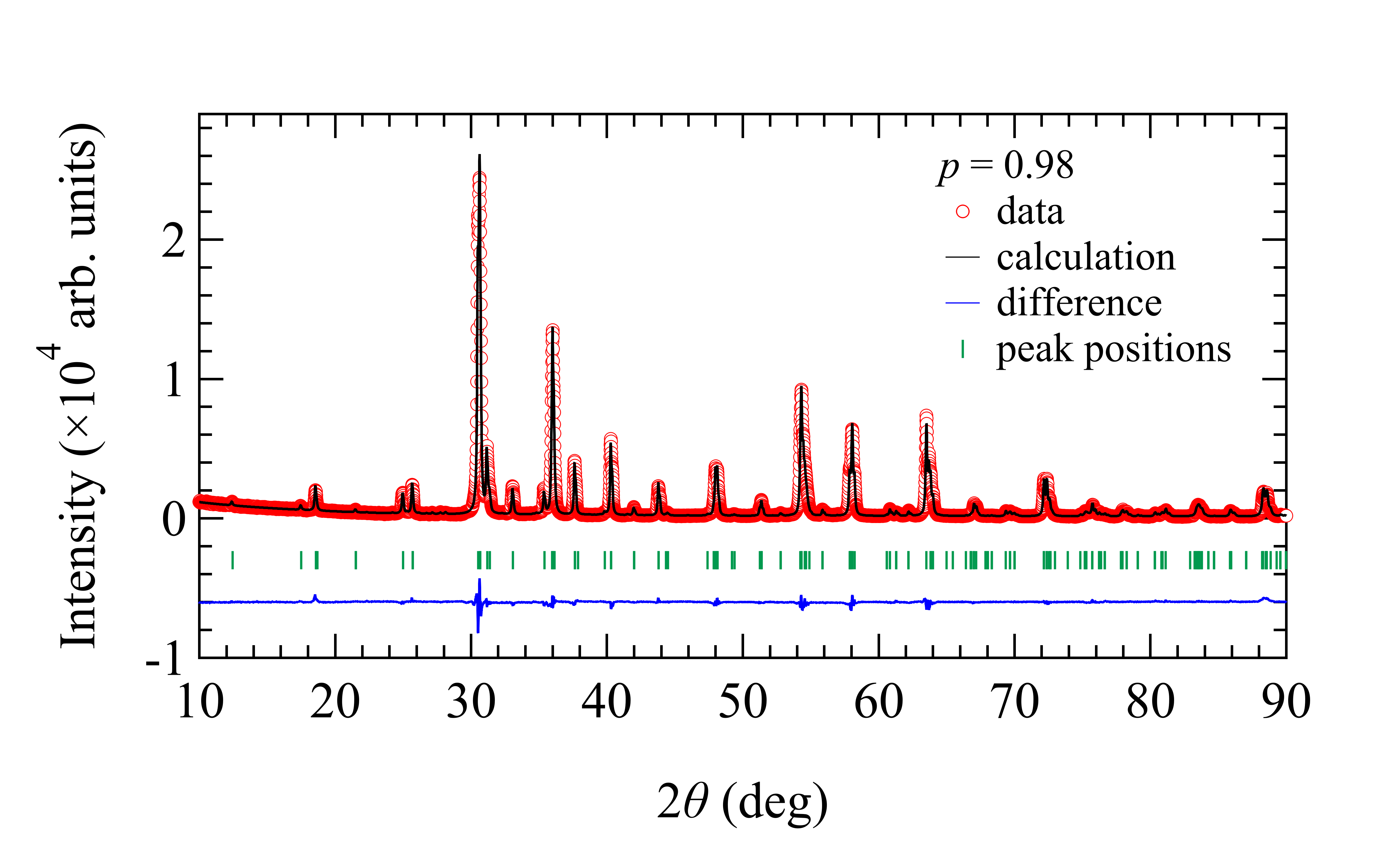}
    \caption{Refinement result of the $p=0.98$ sample.}
    \label{fig_S3-11}
\end{figure}

\begin{table}[ht!]
	\centering
	\caption{Refined atomic parameters of the $p=0$ sample. Refined lattice parameters are $a = b = 5.857241(44)$ Å and $c = 14.253215(78)$ Å. $R_\mathrm{wp} = 8.81\%$ and GoF = 2.83.}
	\begin{tabular}{ccccc}
  	\toprule
 	 Atom & $x/a$ & $y/b$ & $z/c$  & occupancy  \\
  	\midrule
  	Ba ($2d$) & 1/3 & 2/3 & 0.4256(1) & 1.00 \\
  	Sn ($2d$) & 1/3 & 2/3 & 0.6822(1) & 1.00 \\
  	Zn ($2d$) & 1/3 & 2/3 & 0.955(43) & 0.50 \\
 	Ga ($2d$) & 1/3 & 2/3 & 0.955(41) & 0.50 \\
  	Ga ($2c$) & 0   & 0   & 0.3731(2) & 1.00 \\
   	Ga ($1a$) & 0   & 0   & 0 		  & 1.00 \\
  	Ga ($6i$) & 0.1699(2) & -0.1699(2) & 0.17031(8) & 1.00 \\
  	O  ($2c$) & 0   & 0   & 0.2373(9) & 1.00 \\
  	O  ($2d$) & 1/3 & 2/3 & 0.0968(8) & 1.00 \\
  	O  ($6i$) & 0.1572(7) & -0.1572(7) & 0.9144(5) & 1.00 \\
  	O  ($6i$) & 0.4946(9) & -0.4946(9) & 0.2375(6) & 1.00 \\
  	O  ($6i$) & 0.1711(6) & -0.1711(6) & 0.5892(5) & 1.00 \\
  	\bottomrule
	\end{tabular}
	\label{structure_0p00}
\end{table}

\begin{table}[ht!]
	\centering
	\caption{Refined atomic parameters of the $p=0.44$ sample. Refined lattice parameters are $a = b = 5.85692$ Å and $c = 14.25790$ Å. $R_\mathrm{wp} = 5.16\%$ and GoF = 1.52.}
	\begin{tabular}{ccccc}
  	\toprule
 	 Atom & $x/a$ & $y/b$ & $z/c$  & occupancy  \\
  	\midrule
  	Ba ($2d$) & 1/3 & 2/3 & 0.42463 & 1.00  \\
  	Sn ($2d$) & 1/3 & 2/3 & 0.68137 & 1.00 \\
  	Zn ($2d$) & 1/3 & 2/3 & 0.9548 & 0.50 \\
 	Ga ($2d$) & 1/3 & 2/3 & 0.9548 & 0.50 \\
  	Ga ($2c$) & 0   & 0   & 0.3721 & 1.00 \\
	Cr ($1a$) & 0   & 0   & 0 	      & 0.445(14)	 \\
   	Ga ($1a$) & 0   & 0   & 0 		  & 0.555(14)  \\
	Cr ($6i$) & 0.1690 & -0.1690 & 0.17114 & 0.441(4) \\
  	Ga ($6i$) & 0.1690 & -0.1690 & 0.17114 & 0.559(4) \\
  	O  ($2c$) & 0   & 0   & 0.2380 & 1.00 \\
  	O  ($2d$) & 1/3 & 2/3 & 0.0913 & 1.00 \\
  	O  ($6i$) & 0.1527 & -0.1527 & 0.9103 & 1.00 \\
  	O  ($6i$) & 0.4920 & -0.4920 & 0.2403 & 1.00 \\
  	O  ($6i$) & 0.1745 & -0.1745 & 0.5925 & 1.00 \\
  	\bottomrule
	\end{tabular}
	\label{structure_0p43}
\end{table}

\begin{table}[ht!]
	\centering
	\caption{Refined atomic parameters of the $p=0.51$ sample. Refined lattice parameters are $a = b = 5.85382$ Å and $c = 14.24923$ Å. $R_\mathrm{wp} = 6.49\%$ and GoF = 1.76.}
	\begin{tabular}{ccccc}
  	\toprule
 	 Atom & $x/a$ & $y/b$ & $z/c$  & occupancy  \\
  	\midrule
  	Ba ($2d$) & 1/3 & 2/3 & 0.4251 & 1.00  \\
  	Sn ($2d$) & 1/3 & 2/3 & 0.6812 & 1.00 \\
  	Zn ($2d$) & 1/3 & 2/3 & 0.9553 & 0.50 \\
 	Ga ($2d$) & 1/3 & 2/3 & 0.9553 & 0.50 \\
  	Ga ($2c$) & 0   & 0   & 0.3712 & 1.00 \\
  	Cr ($1a$) & 0   & 0   & 0 	      & 0.409(18)	 \\
  	Ga ($1a$) & 0   & 0   & 0 		  & 0.591(18)  \\
  	Cr ($6i$) & 0.1689 & -0.1689 & 0.17114 & 0.523(5) \\
  	Ga ($6i$) & 0.1689 & -0.1689 & 0.17114 & 0.477(5) \\
  	O  ($2c$) & 0   & 0   & 0.2476 & 1.00 \\
  	O  ($2d$) & 1/3 & 2/3 & 0.0888 & 1.00 \\
  	O  ($6i$) & 0.1592 & -0.1592 & 0.9120 & 1.00 \\
  	O  ($6i$) & 0.4899 & -0.4899 & 0.2407 & 1.00 \\
  	O  ($6i$) & 0.1679 & -0.1679 & 0.5878 & 1.00 \\
  	\bottomrule
	\end{tabular}
	\label{structure_0p51}
\end{table}

\begin{table}[ht!]
	\centering
	\caption{Refined atomic parameters of the $p=0.60$ sample. Refined lattice parameters are $a = b = 5.85502$ Å and $c = 14.25203$ Å. $R_\mathrm{wp} = 6.05\%$ and GoF = 1.68.}
	\begin{tabular}{ccccc}
  	\toprule
 	 Atom & $x/a$ & $y/b$ & $z/c$  & occupancy  \\
  	\midrule
  	Ba ($2d$) & 1/3 & 2/3 & 0.42493 & 1.00  \\
  	Sn ($2d$) & 1/3 & 2/3 & 0.68166 & 1.00 \\
  	Zn ($2d$) & 1/3 & 2/3 & 0.9552 & 0.50 \\
 	Ga ($2d$) & 1/3 & 2/3 & 0.9552 & 0.50 \\
  	Ga ($2c$) & 0   & 0   & 0.3717 & 1.00 \\
  	Cr ($1a$) & 0   & 0   & 0 	      & 0.599(15)	 \\
  	Ga ($1a$) & 0   & 0   & 0 		  & 0.401(15)  \\
  	Cr ($6i$) & 0.1691 & -0.1691 & 0.17067 & 0.604(4) \\
  	Ga ($6i$) & 0.1691 & -0.1691 & 0.17067 & 0.396(4) \\
  	O  ($2c$) & 0   & 0   & 0.2448 & 1.00 \\
  	O  ($2d$) & 1/3 & 2/3 & 0.0913 & 1.00 \\
  	O  ($6i$) & 0.1566 & -0.1566 & 0.9122 & 1.00 \\
  	O  ($6i$) & 0.4927 & -0.4927 & 0.2400 & 1.00 \\
  	O  ($6i$) & 0.1693 & -0.1693 & 0.5889 & 1.00 \\
  	\bottomrule
	\end{tabular}
	\label{structure_0p60}
\end{table}

\begin{table}[ht!]
	\centering
	\caption{Refined atomic parameters of the $p=0.63$ sample. Refined lattice parameters are $a = b = 5.85676$ Å and $c = 14.25397$ Å. $R_\mathrm{wp} = 5.06\%$ and GoF = 1.52.}
	\begin{tabular}{ccccc}
  	\toprule
 	 Atom & $x/a$ & $y/b$ & $z/c$  & occupancy  \\
  	\midrule
  	Ba ($2d$) & 1/3 & 2/3 & 0.42444 & 1.00  \\
  	Sn ($2d$) & 1/3 & 2/3 & 0.6815 & 1.00 \\
  	Zn ($2d$) & 1/3 & 2/3 & 0.9548 & 0.50 \\
 	Ga ($2d$) & 1/3 & 2/3 & 0.9548 & 0.50 \\
  	Ga ($2c$) & 0   & 0   & 0.3732 & 1.00 \\
  	Cr ($1a$) & 0   & 0   & 0 	      & 0.625(16)	 \\
  	Ga ($1a$) & 0   & 0   & 0 		  & 0.375(16)  \\
  	Cr ($6i$) & 0.1687 & -0.1687 & 0.17032 & 0.631(4) \\
  	Ga ($6i$) & 0.1687 & -0.1687 & 0.17032 & 0.369(4) \\
  	O  ($2c$) & 0   & 0   & 0.2390 & 1.00 \\
  	O  ($2d$) & 1/3 & 2/3 & 0.0945 & 1.00 \\
  	O  ($6i$) & 0.1545 & -0.1545 & 0.9123 & 1.00 \\
  	O  ($6i$) & 0.4923 & -0.4923 & 0.2367 & 1.00 \\
  	O  ($6i$) & 0.1754 & -0.1754 & 0.59183 & 1.00 \\
  	\bottomrule
	\end{tabular}
	\label{structure_0p63}
\end{table}

\begin{table}[ht!]
	\centering
	\caption{Refined atomic parameters of the $p=0.67$ sample. Refined lattice parameters are $a = b = 5.85758$ Å and $c = 14.25553$ Å. $R_\mathrm{wp} = 5.45\%$ and GoF = 1.61.}
	\begin{tabular}{ccccc}
  	\toprule
 	 Atom & $x/a$ & $y/b$ & $z/c$  & occupancy  \\
  	\midrule
  	Ba ($2d$) & 1/3 & 2/3 & 0.4240 & 1.00  \\
  	Sn ($2d$) & 1/3 & 2/3 & 0.6814 & 1.00 \\
  	Zn ($2d$) & 1/3 & 2/3 & 0.9546 & 0.50 \\
 	Ga ($2d$) & 1/3 & 2/3 & 0.9546 & 0.50 \\
  	Ga ($2c$) & 0   & 0   & 0.3732 & 1.00 \\
  	Cr ($1a$) & 0   & 0   & 0 	      & 0.621(17)	 \\
  	Ga ($1a$) & 0   & 0   & 0 		  & 0.379(17)  \\
  	Cr ($6i$) & 0.1689 & -0.1689 & 0.17019 & 0.672(4) \\
  	Ga ($6i$) & 0.1689 & -0.1689 & 0.17019 & 0.328(4) \\
  	O  ($2c$) & 0   & 0   & 0.2358 & 1.00 \\
  	O  ($2d$) & 1/3 & 2/3 & 0.0966 & 1.00 \\
  	O  ($6i$) & 0.1558 & -0.1558 & 0.9119 & 1.00 \\
  	O  ($6i$) & 0.4925 & -0.4925 & 0.2363 & 1.00 \\
  	O  ($6i$) & 0.1740 & -0.1740 & 0.5935 & 1.00 \\
  	\bottomrule
	\end{tabular}
	\label{structure_0p67}
\end{table}

\begin{table}[ht!]
	\centering
	\caption{Refined atomic parameters of the $p=0.71$ sample. Refined lattice parameters are $a = b = 5.85367$ Å and $c = 14.24537$ Å. $R_\mathrm{wp} = 6.14\%$ and GoF = 1.69.}
	\begin{tabular}{ccccc}
  	\toprule
 	 Atom & $x/a$ & $y/b$ & $z/c$  & occupancy  \\
  	\midrule
  	Ba ($2d$) & 1/3 & 2/3 & 0.4245 & 1.00  \\
  	Sn ($2d$) & 1/3 & 2/3 & 0.6817 & 1.00 \\
  	Zn ($2d$) & 1/3 & 2/3 & 0.9551 & 0.50 \\
 	Ga ($2d$) & 1/3 & 2/3 & 0.9551 & 0.50 \\
  	Ga ($2c$) & 0   & 0   & 0.3722 & 1.00 \\
  	Cr ($1a$) & 0   & 0   & 0 	      & 0.703(16)	 \\
  	Ga ($1a$) & 0   & 0   & 0 		  & 0.297(16)  \\
  	Cr ($6i$) & 0.1690 & -0.1690 & 0.17036 & 0.715(4) \\
  	Ga ($6i$) & 0.1690 & -0.1690 & 0.17036 & 0.285(4) \\
  	O  ($2c$) & 0   & 0   & 0.2425 & 1.00 \\
  	O  ($2d$) & 1/3 & 2/3 & 0.0938 & 1.00 \\
  	O  ($6i$) & 0.1548 & -0.1548 & 0.9126 & 1.00 \\
  	O  ($6i$) & 0.4895 & -0.4895 & 0.2392 & 1.00 \\
  	O  ($6i$) & 0.1732 & -0.1732 & 0.5906 & 1.00 \\
  	\bottomrule
	\end{tabular}
	\label{structure_0p71}
\end{table}

\begin{table}[ht!]
	\centering
	\caption{Refined atomic parameters of the $p=0.83$ sample. Refined lattice parameters are $a = b = 5.85338$ Å and $c = 14.2452$ Å. $R_\mathrm{wp} = 6.24\%$ and GoF = 1.81.}
	\begin{tabular}{ccccc}
  	\toprule
 	 Atom & $x/a$ & $y/b$ & $z/c$  & occupancy  \\
  	\midrule
  	Ba ($2d$) & 1/3 & 2/3 & 0.4247 & 1.00  \\
  	Sn ($2d$) & 1/3 & 2/3 & 0.6817 & 1.00 \\
  	Zn ($2d$) & 1/3 & 2/3 & 0.9550 & 0.50 \\
 	Ga ($2d$) & 1/3 & 2/3 & 0.9550 & 0.50 \\
  	Ga ($2c$) & 0   & 0   & 0.3734 & 1.00 \\
  	Cr ($1a$) & 0   & 0   & 0 	      & 0.823(19)	 \\
  	Ga ($1a$) & 0   & 0   & 0 		  & 0.177(19)  \\
  	Cr ($6i$) & 0.1685 & -0.1685 & 0.1695 & 0.829(5) \\
  	Ga ($6i$) & 0.1685 & -0.1685 & 0.1695 & 0.171(5) \\
  	O  ($2c$) & 0   & 0   & 0.2383 & 1.00 \\
  	O  ($2d$) & 1/3 & 2/3 & 0.0969 & 1.00 \\
  	O  ($6i$) & 0.1560 & -0.1560 & 0.9125 & 1.00 \\
  	O  ($6i$) & 0.4944 & -0.4944 & 0.2381 & 1.00 \\
  	O  ($6i$) & 0.1771 & -0.1771 & 0.5927 & 1.00 \\
  	\bottomrule
	\end{tabular}
	\label{structure_0p83}
\end{table}

\begin{table}[ht!]
	\centering
	\caption{Refined atomic parameters of the $p=0.86$ sample. Refined lattice parameters are $a = b = 5.85326$ Å and $c = 14.2454$ Å. $R_\mathrm{wp} = 5.80\%$ and GoF = 1.69.}
	\begin{tabular}{ccccc}
  	\toprule
 	 Atom & $x/a$ & $y/b$ & $z/c$  & occupancy  \\
  	\midrule
  	Ba ($2d$) & 1/3 & 2/3 & 0.4236 & 1.00  \\
  	Sn ($2d$) & 1/3 & 2/3 & 0.6814 & 1.00 \\
  	Zn ($2d$) & 1/3 & 2/3 & 0.9550 & 0.50 \\
 	Ga ($2d$) & 1/3 & 2/3 & 0.9550 & 0.50 \\
  	Ga ($2c$) & 0   & 0   & 0.3735 & 1.00 \\
  	Cr ($1a$) & 0   & 0   & 0 	      & 0.857(21)	 \\
  	Ga ($1a$) & 0   & 0   & 0 		  & 0.143(21)  \\
  	Cr ($6i$) & 0.1686 & -0.1686 & 0.1695 & 0.861(6) \\
  	Ga ($6i$) & 0.1686 & -0.1686 & 0.1695 & 0.139(6) \\
  	O  ($2c$) & 0   & 0   & 0.2388 & 1.00 \\
  	O  ($2d$) & 1/3 & 2/3 & 0.0961 & 1.00 \\
  	O  ($6i$) & 0.1546 & -0.1546 & 0.9129 & 1.00 \\
  	O  ($6i$) & 0.4931 & -0.4931 & 0.2390 & 1.00 \\
  	O  ($6i$) & 0.1771 & -0.1771 & 0.5945 & 1.00 \\
  	\bottomrule
	\end{tabular}
	\label{structure_0p86}
\end{table}

\begin{table}[ht!]
	\centering
	\caption{Refined atomic parameters of the $p=0.93$ sample. Refined lattice parameters are $a = b = 5.85457$ Å and $c = 14.2491$ Å. $R_\mathrm{wp} = 6.38\%$ and GoF = 1.82.}
	\begin{tabular}{ccccc}
  	\toprule
 	 Atom & $x/a$ & $y/b$ & $z/c$  & occupancy  \\
  	\midrule
  	Ba ($2d$) & 1/3 & 2/3 & 0.4237 & 1.00  \\
  	Sn ($2d$) & 1/3 & 2/3 & 0.6824 & 1.00 \\
  	Zn ($2d$) & 1/3 & 2/3 & 0.9554 & 0.50 \\
 	Ga ($2d$) & 1/3 & 2/3 & 0.9554 & 0.50 \\
  	Ga ($2c$) & 0   & 0   & 0.3734 & 1.00 \\
  	Cr ($1a$) & 0   & 0   & 0 	      & 0.937(25)	 \\
  	Ga ($1a$) & 0   & 0   & 0 		  & 0.063(25)  \\
  	Cr ($6i$) & 0.1677 & -0.1677 & 0.1686 & 0.933(7) \\
  	Ga ($6i$) & 0.1677 & -0.1677 & 0.1686 & 0.067(7) \\
  	O  ($2c$) & 0   & 0   & 0.245 & 1.00 \\
  	O  ($2d$) & 1/3 & 2/3 & 0.1021 & 1.00 \\
  	O  ($6i$) & 0.156 & -0.156 & 0.9126 & 1.00 \\
  	O  ($6i$) & 0.4889 & -0.4889 & 0.2355 & 1.00 \\
  	O  ($6i$) & 0.1769 & -0.1769 & 0.5878 & 1.00 \\
  	\bottomrule
	\end{tabular}
	\label{structure_0p93}
\end{table}

\begin{table}[ht!]
	\centering
	\caption{Refined atomic parameters of the $p=0.98$ sample. Refined lattice parameters are $a = b = 5.85204$ Å and $c = 14.2434$ Å. $R_\mathrm{wp} = 8.30\%$ and GoF = 2.18.}
	\begin{tabular}{ccccc}
  	\toprule
 	 Atom & $x/a$ & $y/b$ & $z/c$  & occupancy  \\
  	\midrule
  	Ba ($2d$) & 1/3 & 2/3 & 0.4237 & 1.00  \\
  	Sn ($2d$) & 1/3 & 2/3 & 0.6818 & 1.00   \\
  	Zn ($2d$) & 1/3 & 2/3 & 0.9560 & 0.50  \\
 	Ga ($2d$) & 1/3 & 2/3 & 0.9560 & 0.50 \\
  	Ga ($2c$) & 0   & 0   & 0.3724 & 1.00 \\
  	Cr ($1a$) & 0   & 0   & 0 	      & 1.00(2)  \\
  	Ga ($1a$) & 0   & 0   & 0 		  & 0.00(2)  \\
  	Cr ($6i$) & 0.1688 & -0.1688 & 0.1707 & 0.976(4) \\
  	Ga ($6i$) & 0.1688 & -0.1688 & 0.1707 & 0.024(4) \\
  	O  ($2c$) & 0   & 0   & 0.244   & 1.00 \\
  	O  ($2d$) & 1/3 & 2/3 & 0.0929  & 1.00 \\
  	O  ($6i$) & 0.1555 & -0.1555 & 0.9126 & 1.00 \\
  	O  ($6i$) & 0.4853 & -0.4853 & 0.2387 & 1.00 \\
  	O  ($6i$) & 0.1709 & -0.1709 & 0.5895 & 1.00 \\
  	\bottomrule
	\end{tabular}
	\label{structure_0p98}
\end{table}

\end{document}